# ARTIFICIAL INTELLIGENCE, HUMAN RIGHTS, DEMOCRACY, AND THE RULE OF LAW

## A PRIMER

PREPARED TO SUPPORT THE **FEASIBILITY STUDY** PUBLISHED BY THE COUNCIL OF EUROPE'S AD HOC COMMITTEE ON ARTIFICIAL INTELLIGENCE

DAVID LESLIE, CHRISTOPHER BURR, MHAIRI AITKEN, JOSH COWLS, MIKE KATELL, & MORGAN BRIGGS

With a foreword by

LORD TIM CLEMENT-JONES

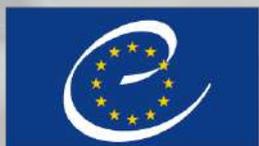

The Public Policy Programme at The Alan Turing Institute was set up in May 2018 with the aim of developing research, tools, and techniques that help governments innovate with data-intensive technologies and improve the quality of people's lives. We work alongside policy makers to explore how data science and artificial intelligence can inform public policy and improve the provision of public services. We believe that governments can reap the benefits of these technologies only if they make considerations of ethics and safety a first priority.

Please note that this primer is a living document that will evolve and improve with input from users, affected stakeholders, and interested parties. We need your participation. Please share feedback with us at policy@turing.ac.uk  This research was supported, in part, by a grant from ESRC (ES/T007354/1) and from the public funds that make the Turing's Public Policy Programme possible. https://www.turing.ac.uk/research/research-programmes/public-policy

The opinions expressed in this work are the responsibility of the authors and do not necessarily reflect the official policy of the Council of Europe. The content on which much of this primer was built was taken from *The Feasibility Study* published by the Ad Hoc Committee on Artificial Intelligence published in December 2020. Readers are recommended to refer directly to this document for expansion on the ideas contained herein: https://rm.coe.int/cahai-2020-23-final-eng-feasibility-study-/1680a0c6da

# TABLE OF CONTENTS

# FOREWORD

It has never been clearer, particularly after this year of COVID has exposed our ever greater reliance on digital technology, that we need to retain public trust in the adoption of AI.

To do that we need, whilst realising the opportunities, to mitigate the risks involved in the application of AI. This brings with it the need for a clear standard of accountability and ethical behaviour.

If 2019 was the year when countries signed up to internationally agreed AI ethical principles such as those in the OECD Recommendation on AI, and the G20 non-binding principles on AI, 2020 was the year when the international AI community started to move towards deciding how to instill them in the development and deployment of AI systems.

Making ethical AI a reality involves assessing the risks of AI in context particularly in terms of impact on civil and social rights and then, depending on the risk assessed, setting standards or regulating for the ethical design, development and deployment of AI systems.

A key initiative in that process has been the Feasibility Study drawn up and agreed in December by the Council of Europe's Ad Hoc Committee on Artificial Intelligence (CAHAI) which explores options for an international legal response based on Council of Europe standards in the field of human rights, democracy, and the rule of law.

The key question is whether there are responses to the specific risks and opportunities presented by AI systems which can and should be met by the use of binding and non-binding international legal instruments through the agency of the Council of Europe which is the custodian of the European Convention on Human Rights, Convention 108+, which safeguards the processing of personal data, and the European Social Charter.

Now that the Council and CAHAI are entering the stakeholder consultation phase for the Feasibility Study, it is crucial, if its potential is to be realised, and the right choices are to be made particularly in terms of legal instrument and oversight and compliance mechanisms, that the societal and regulatory implications of its principles-based proposals and approach are fully understood.

This superb Primer produced by The Alan Turing Institute as a companion to the Feasibility Study and designed to explain its context and assist with the consultation, is a model of clarity. It will undoubtedly increase public engagement and ensure that a wide, and, at the same time, informed, debate can take place. This is a vital area of public policy where broad informed discussion by the many, particularly on the values to be adopted is crucial. This Primer will ensure that it is not just left to the decision making of the specialist few.

*Lord Tim Clement-Jones*
London, 2021

# 01 INTRODUCTION

## THE PURPOSE OF THIS PRIMER

It is a remarkable fact that rapid advancements in artificial intelligence (AI) and data-driven technologies over the last two decades have placed contemporary society at a pivot-point in deciding what shape the future of humanity will take. On the one hand, the flourishing of societally beneficial AI innovation promises, among other things, to help us tackle climate change and biodiversity loss; to equitably improve medical care, living standards, transportation, and agricultural production; and to address many of the social injustices and material inequalities that beset today's world. On the other hand, the proliferation of irresponsible AI innovations is revealing warning signs of the potential troubles that may lie ahead if the advancement of these technologies continues on its current worrying trajectory.

We see these warning signs, for instance, in the growing risks to cherished rights to privacy, self-expression, association, and consent, as well as to other civil liberties and social freedoms, that digital surveillance infrastructures like live facial recognition now increasingly pose. We see them in the transformative effects already apparent in the broad-scale proliferation of individual-targeting algorithmic curation and data-driven behavioural manipulation which have bolstered the revenues of Big Tech platforms all while fostering global crises of social distrust, contagions of disinformation, and heightening levels of cultural and political polarisation. We see them too in the way that the application of predictive risk models and algorithmically-enhanced digital tracking capacities in high impact areas like law enforcement has functioned to reinforce and further entrench patterns of structural discrimination, systemic marginalisation, and inequality.

Recognising the need for democratically-led human intervention in setting AI innovation on the right track, the Council of Europe's Committee of Ministers adopted the terms of reference, in September 2019, for the Ad Hoc Committee on Artificial Intelligence (CAHAI). The CAHAI is charged with examining the feasibility and potential elements of a legal framework for the design, development, and deployment of AI systems that accord with Council of Europe standards across the interrelated areas of human rights, democracy, and the rule of law.

As a first and necessary step in carrying out this responsibility, the CAHAI's *Feasibility Study*, adopted by its plenary in December 2020, has explored options for an international legal response that fills existing gaps in legislation and tailors the use of binding and non-binding legal instruments to the specific risks and opportunities presented by AI systems. The *Study* examines how the fundamental rights and freedoms that are already codified in international human rights law can be used as the basis for such a legal framework. It proposes nine principles and priorities that are fitted to the novel challenges posed by the design, development, and deployment of AI systems. When codified into law, these principles and priorities create a set of interlocking rights and obligations that will work towards ensuring that the design and use of AI technologies conform to the values of human rights, democracy, and the rule of law. *The Feasibility Study* concludes that current rules and legal regimes are neither adequate for safeguarding these basic values as they pertain to AI, nor suitable, in and of themselves, for creating an AI innovation environment that can be deemed sufficiently trustworthy for steering AI and data-intensive technologies in the right direction. A new legal framework is needed.

The purpose of this primer is to introduce the main concepts and principles presented in the CAHAI's *Feasibility Study* for a general, non-technical audience. It also aims to provide some background information on the areas of AI innovation, human rights law, technology policy, and compliance mechanisms covered therein. In keeping with the Council of Europe's commitment to broad multistakeholder consultations, outreach, and engagement, this primer has been designed to help facilitate the meaningful and informed participation of an inclusive group of stakeholders as the CAHAI seeks feedback and guidance regarding the essential issues raised by *The Feasibility Study*.

# HOW TO USE THIS PRIMER

This primer has been designed to support both readers who have no technical background and readers who may have some but are still interested in "brushing up" on one or a few of the topics that are covered by *The Feasibility Study*. For this reason, we have written the chapters in a modular fashion, meaning that the reader is welcomed to either select those topics and sections that are of most interest (and focus on them) or to engage with the primer from start to finish.

The first three chapters provide stage-setting information about AI and machine learning technologies (Ch. 2); human rights, democracy, and the rule of law (Ch. 3); and the risks and opportunities presented by AI systems in the human rights context (Ch. 4). The primer then moves on to discuss some of the more specific subjects covered in *The Feasibility Study*. Chapter 5 lays out the nine principles and priorities that have been proposed by the CAHAI as an anchor for a values-based and cross-sectoral legal framework. It then presents the points of contact between these principles and priorities and the key rights and obligations that will allow them to be translated into statute. Chapter 6 provides a summary of the landscape of legal instruments that may be integrated into a larger arrangement of binding and non-binding legal mechanisms. Finally, Chapter 7 presents the spectrum of compliance tools that are available to support, operationalise, and underwrite the constraints set in place by a legal framework.

At the very end of this primer, you will find a glossary of relevant terms and an annotated list of publications, which includes some of the previous work done by the Council of Europe and others in the field of AI standards and regulation and adjacent areas of technology policy.

Because there is no substitute for the great accomplishment of the original, we highly recommend that readers directly engage with *The Feasibility Study* itself and use this primer merely as a companion, ready-to-hand for contextual information, clarification, and condensed presentation.

# 02  HOW DO AI SYSTEMS WORK?

Before launching into an exploration of how a framework of binding and non-binding legal instruments can align the design, development, and deployment of AI technologies with human rights, democracy, and the rule of law, we now present an explainer of the basic technical concepts, the types of machine learning, and the stages of the AI lifecycle.

## TECHNICAL CONCEPTS

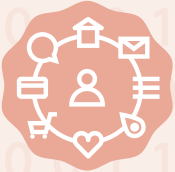

### PERSONAL DATA

Data that can be used to identify an individual. Examples of personal data may include things such as first name and surname, address, location data, forms of identification (e.g. passport, national ID), amongst others.

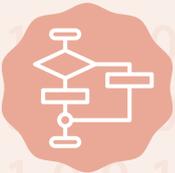

### ALGORITHM

A computational process or set of rules that are performed to solve some problem. A computer is typically used to carry out complex algorithms, but a human could also follow an algorithmic process, such as by following a recipe or using a mathematical formula to solve an equation.

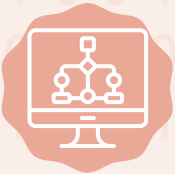

### MACHINE LEARNING (ML)

A type of computing used to find patterns in data and to make predictions of an outcome for a particular instance. "Learning" is a bit misleading, as the computer does not learn in the same way as humans do. Instead, the computer is able to find similarities and differences in the data through the repetitious tuning of its parameters (often called "training"). When the input data changes, the outputs also change accordingly, meaning the computer learns to detect new patterns. This is accomplished by applying a mathematical formula to large amounts of input data to produce a corresponding outcome. This is described in more detail in the next section.

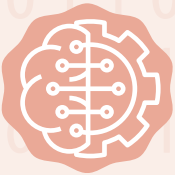

### ARTIFICIAL INTELLIGENCE (AI)

There are many ways that AI has been defined over the last several decades, but for the purposes of this primer, we will stick to defining it by describing what it does, i.e. what role it plays in the human world: AI systems are algorithmic models that carry out cognitive or perceptual functions in the world that were previously reserved for thinking, judging, and reasoning human beings.

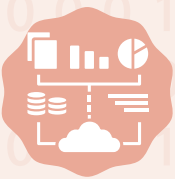

### BIG DATA

Datasets that are voluminous, often require large amounts of storage, and contain vast amounts of quantitative data that can be used for revealing patterns or trends. Data contained within these large datasets can range in type (e.g. numbers, words, images) and be either specific to a purpose and tabular (structured) or general and varied (unstructured).

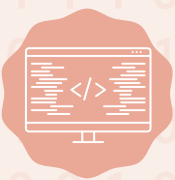

### DATA SCIENCE

A field that includes elements from various disciplines including computer science, mathematics, statistics, and the social sciences, and is generally focused on extracting insights and patterns from datasets to answer or address a specific question or problem.

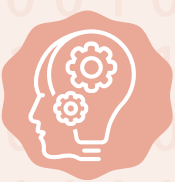

### INTERPRETABILITY

If a human is able to identify how an AI or machine learning system came to some decision, or explain why it behaved in some way, then the system can be described as interpretable. Interpretability may also refer to the transparency of the processes by which the system was developed.

# TYPES OF MACHINE LEARNING

## SUPERVISED LEARNING

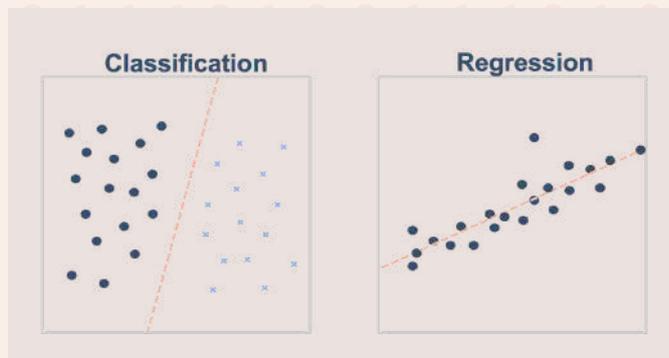

Supervised learning models are trained on datasets that contain labelled data. "Learning" occurs in these models when numerous examples are used to train an algorithm to map input variables (often called features) onto desired outputs (also called target variables or labels). On the basis of these examples, ML models become capable of identifying patterns that link inputs to outputs. Such ML models can then reproduce these patterns by employing the rules honed during training to transform new inputs received into classifications or predictions. A classic example of supervised learning is using various variables such as the presence of words like "lottery" or "you won" to predict whether or not an email should be classified as spam or not spam. Supervised learning can take the form of classification such as a prediction that an email is or is not spam, or regression which involves determining the relationship between input variables and a target variable. While linear regression and classification are the simplest forms of supervised learning, other supervised models such as support vector machines and random forests are also common applications.

## UNSUPERVISED LEARNING

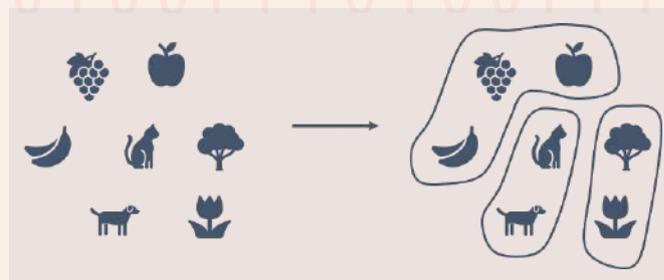

The goal of unsupervised learning is for the system to identify patterns amongst the data, while supervised learning is a process of mapping relationships between data points, as in the comparison of two images where the objects in one have already been identified. Unsupervised learning involves identifying patterns and structures by measuring the densities or similarities of data points in the dataset. A common application of unsupervised learning is clustering, in which the model receives unlabelled input data and determines similarities and differences amongst the input data points, resulting in clusters based on similar traits that are important factors in categorising the input data. In the example above, the model is given types of fruit, animals, a flower, and a tree. Based on traits unique to each of the categories, clustering is able to separate animals, fruits, and plants out into three separate clusters. Dimensionality reduction is another form of unsupervised learning.

## REINFORCEMENT LEARNING

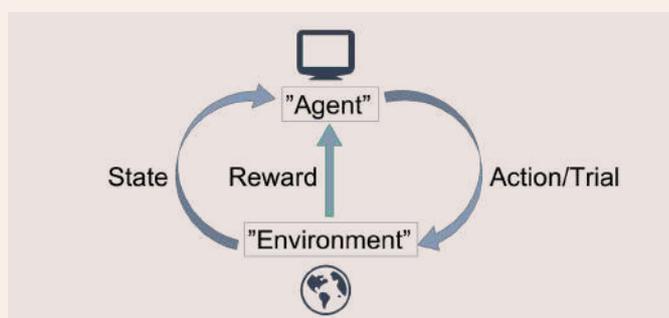

Reinforcement learning models learn on the basis of their interactions with a virtual or real environment rather than existing data. Reinforcement learning "agents" search for an optimal way to complete a task by taking a series of steps that maximise the probability of achieving that task. Depending on the success or failure of the steps they take, they are rewarded or penalised. These "agents" are programmed to choose their steps to maximise their reward. They "learn" from past rewards and failures, improve with multiple iterations of trial and error, and may be designed to develop long-term strategies to maximise their reward overall rather than looking only at their next step. A common example of reinforcement learning can be found in the development of autonomous vehicles (self-driving cars). Reinforcement learning is used to improve the vehicle's performance in a simulated environment, testing for things such as response to traffic controls and acceleration. Through these interactions with the simulated environment, the reinforcement learning "agents" are penalised or rewarded based on task completion, thereby impacting the vehicle's future performance.

## STAGES OF THE AI LIFECYCLE

### DESIGN

### Project Planning

A project team must decide what the project's goals are at the outset. Tasks in this stage may include stakeholder engagement activities, wider impact assessments, mapping of key stages within the project, or an assessment of resources and capabilities within the team or organisation. For example, an AI project team is deciding whether or not to use an AI application within an agricultural setting to predict which fields are likely to be arable over the next five years, and what the possible crop yield will be. This planning allows the project team to reflect on the ethical, socio-economic legal, and technical issues before investing any resources into developing the system.

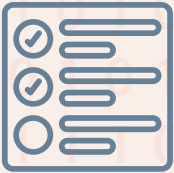

### Problem Formulation

A project team needs to determine what problem their model will address, along with deciding what input data is needed and for what purpose. The team should consider ethical and legal implications of the uses of data and provide a thorough account of intended and unintended consequences of use. For instance, the team has determined the overarching theme of the project will involve crop yields. This more precise formulation helps to identify a specific question that can be approached through data and ensure that the result will accord with ethical and legal considerations, such as biodiversity or land use.

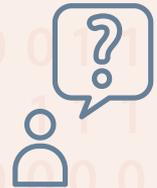

### Data Extraction or Procurement

This stage involves the processes by which data is gathered for the problem at hand. Data extraction may involve web scraping processes or data recording through surveys or similar methodologies, whereas procurement may involve legal agreements to obtain already existing datasets. In our running example, the team has decided their problem will involve determining factors important in predicting crop yields in a given agricultural season. They decide to request data from a government agency and farming co-ops, both of which require legal data sharing agreements.

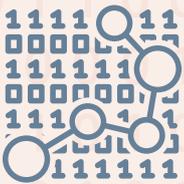

### Data Analysis

At this stage, the project team can begin to inspect the data. Primarily, this will entail a high degree of exploratory data analysis (EDA). EDA involves understanding the makeup of the data through visualisation and summary statistics. Some questions at this stage may include: is there missing data (incomplete data), outliers (unexpected data), unbalanced classes (imbalanced data), or correlation? For example, the team creates visualisations to understand things such as the distribution of crop types across farms, weather conditions, soil pH levels, along with understanding any missing data present.

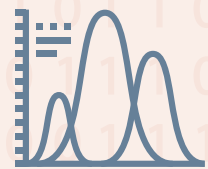

## Preprocessing

The preprocessing stage is often the most time consuming part of the development phase of the AI lifecycle. Preprocessing includes tasks such as data cleaning (reformatting or removing incomplete information), and data wrangling (transforming data into a format conducive for modelling), amongst other processes that feed into the model training process. For example, during preprocessing, the members of the team notice that soil pH levels are treated as both numeric and text string data, which would cause issues when running the model, so they decide to make all of the soil pH levels the same data type by transforming the text string data into numeric data.

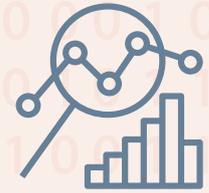

## Model Selection and Training

Models should be selected to serve the problem determined in the design phase. Model types vary in complexity; however, model selection considers other factors such as data types, quantity, and availability. Models that lack sufficient complexity run the risk of underfitting (or failing to account for) the data. Preprocessed data is split into training and testing sets to avoid overfitting. Overfitting occurs when the model reflects the training data too closely and is unable to fit new, "unseen" data to make accurate predictions for inputs that were not in the training set. Training data are used to hone the parameters of the selected model. As an example of model selection, the project team has decided to employ a linear regression model to use past data to predict future crop yields. They wanted a model that was interpretable in order to fully explain the results, so choosing a simple technique like linear regression made sense.

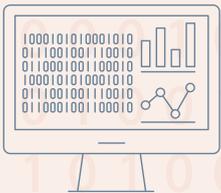

## Model Testing and Validation

After training, the model is then tuned and tested against "unseen" data. Validation sets are used to adjust higher-level aspects of the model (like hyperparameters that govern the way the model learns) and are often created by initially splitting the dataset into three parts, for instance, 60% training data, 20% testing data, and 20% validation data. During validation, elements of the model's architecture can be altered to affect model performance. For instance, the team runs the model and realises the number of variables included is causing overfitting. So, they decide to add a regularisation term (a method used to reduce the error of the model) in order to remove unimportant variables. The model is then tested on unfamiliar data to mimic real world application and to confirm performance and accuracy.

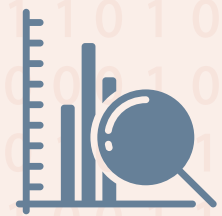

## Model Reporting

After the team trains, validates, and tests the model, model evaluation (including a variety of performance measures and impact assessments), along with detailed information about the model workflow should be produced to better support transparent discussions about the model's output. For example, to complete the development phase, the team documents various performance metrics of their model, along with the processes to get to the current iteration of the model including preprocessing and the decision to add regularisation in the model testing and validation stage.

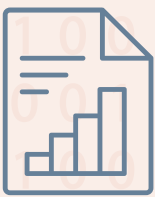

# DEPLOYMENT

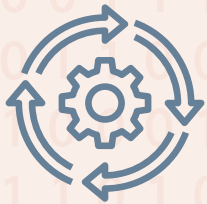

## Model Implementation

The next stage of the AI lifecycle involves deploying the trained model in the real world. Effective implementation allows the model to be incorporated into a larger system. New data is processed by the implemented model to serve the intended purpose determined in the design phase. For instance, the AI project team has decided that the crop yield model is ready to be used. They choose to make it available to several farming co-ops and ask them to run it on their data to see if it provides useful insights.

## User Training

Implementers of the system must be trained to understand the logic of the system, be able to explain its decisions in plain language to decision subjects, and use independent and unbiased judgement to gauge the quality, reliability, and fairness of its outputs. For example, after the team has trained specific users in the agricultural industry on how to use their model, these users will report back on whether they find the system to be useful, reliable, and accurate, amongst other metrics.

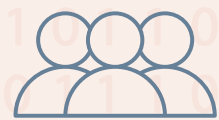

## Monitoring

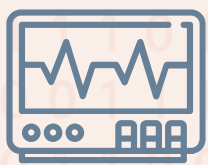

After the model is implemented by the team, it must be monitored to ensure that it is still serving the desired purpose, being used responsibly and within the intended scope, and is responsive to emergent, real-world conditions. For instance, the team notices that a new variable to measure water quality was released by a standards agency. This could cause a lack of standardisation across the data, as it was not an original variable included in the training data set. They decide to incorporate this change into the model to stay current with agriculture norms and practices.

## Updating or Deprovisioning

Over time, the model may lose efficacy, requiring the supervising team to revisit earlier stages of the development phase including model selection and training. If more significant changes are required, the system may need to be deprovisioned, thereby restarting at the design process with project planning. For example, the team has had to retrain the model several times based on new variables and non-standardised data sets. They continue to monitor the model while considering alternative options, including the development of a new system.

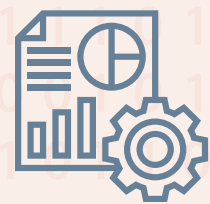

# 03 A BRIEF INTRODUCTION TO HUMAN RIGHTS, DEMOCRACY, AND THE RULE OF LAW

*"All human rights are universal, indivisible and interdependent and interrelated"*

-United Nations Vienna Declaration, 1993

Human rights, democracy, and the rule of law are closely linked. The capacity of legitimate governments to effectively safeguard human rights is predicated on the interdependence of robust and accountable democratic institutions, inclusive and transparent mechanisms of decision-making, and an independent and impartial judiciary that secures the rule of law. Most generally, human rights are the basic rights and freedoms that are possessed by every person in the world from cradle to grave and that preserve and protect the inviolable dignity of each individual regardless of their race, ethnicity, gender, age, sexual orientation, class, religion, disability status, language, nationality, or any other ascribed characteristic. These fundamental rights and freedoms create obligations that bind governments to respecting, protecting, and fulfilling human rights. In the absence of the fulfilment of these duties, individuals are entitled to legal remedies that allow for the redress of any human rights violations.

## HUMAN RIGHTS AT A GLANCE

Historically, the set of basic rights and principles that have come to be known as human rights first emerged in the mid-20th century in the wake of the atrocities and trauma of World War II.

**1948** — The United Nations adopts the **The Universal Declaration of Human Rights** (UDHR), which provides a first international standard for fundamental rights and freedoms. Though not legally binding, this document would become the basis for the many treaties, conventions, and charters on human rights that have been adopted worldwide up to the present.

**1953** — **The European Convention on Human Rights** (ECHR) goes into effect. First drafted by the Council of Europe in 1950, this international treaty enshrines the civil and political rights to which the 47 Member States of the Council are legally bound. Beyond establishing basic rights aimed to safeguard the inviolable dignity of every person, the ECHR placed obligations on governments to protect ordinary people against human rights violations.

**1961** — The Council of Europe releases its **European Social Charter** (ESC) for signatures. This treaty extends basic rights to include social and economic rights covering health, working conditions, housing, migrant labour, gender equality, and social security. Additional protocols were added in 1988 that strengthened equality of opportunity in the workplace, worker participation, and protection of the poor and elderly. A revised ESC was then adopted in 1996.

**1966** — The UN adopts its **International Covenant on Civil and Political Rights** (ICCPR) and **International Covenant on Economic, Social and Cultural Rights** (ICESCR). The ICCPR includes freedom from torture, right to a fair trial, non-discrimination, and privacy rights. The ICESCR extends basic rights to include rights to just working conditions, health, living standards, education, and social security. Taken together, the UN's UDHR, ICCPR and ICESCR are now known as **The International Bill of Human Rights**.

**2009** — **Charter of Fundamental Rights of the European Union** (CFR) goes into full legal force through the Treaty of Lisbon. This codified a basic set of civil, political, social, economic, and cultural rights for citizens of the European Union in EU law. The areas of human rights covered by the CFR include those pertaining to human dignity, fundamental freedoms, equality, solidarity, and economic rights, and rights to participation in the life of the community.

# TWO FAMILIES OF HUMAN RIGHTS

The body of principles that constitutes human rights can be broken down into two groupings:

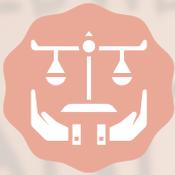
## Civil and Political Rights

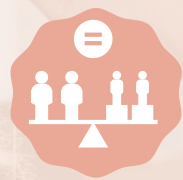
## Social, Economic, and Cultural Rights

Key rights:

- Right to life and human dignity
- Right to physical and mental integrity
- Right to liberty and security of persons
- Freedom from torture and cruel treatment
- Right to a fair trial and due judicial process
- Right to effective remedy
- Freedom of thought, conscience, and religion
- Freedom of expression and opinion
- Right to respect for private and family life
- Right to the protection of personal data
- Right to non-discrimination
- Right to equality before the law
- Freedom of assembly and association
- Right to participate in the conduct of public affairs

Key rights:

- Right to just, safe, and healthy working conditions
- Right to fair remuneration
- Right to vocational training
- Right to equality of opportunity in the workplace
- Right to organise and collectively bargain
- Right to social security
- Right to education
- Right to an adequate standard of living
- Right to social and medical assistance
- Right to the protection of health
- Right of protection for migrant workers
- Right for elderly persons to social protection
- Right to protection against sexual harassment
- Right to protection against poverty and social exclusion

# INTERDEPENDENCE OF HUMAN RIGHTS, DEMOCRACY, AND THE RULE OF LAW

The interdependence of human rights, democracy, and the rule of law originates in their nested and symbiotic character. The legitimacy of democratic institutions is rooted in the notion that each and every citizen is equally entitled to participate in the shared life of the community and in the steering of the collective decisions that impact them. Yet, for citizens to exercise this right to participate in the conduct of public affairs, they must first possess many other interrelated civil, political, social, cultural, and economic rights:

- They must enjoy freedom of thought, association, assembly, and expression.
- They must be granted equal respect before the law and protection from any forms of discrimination that would encumber their full and equitable participation in community life.
- They must have access to the material means of participation through the provision of proper education, adequate living and working standards, health, safety, and social security.
- They must be able to access effective judicial remedies in the event that any of their basic rights are harmed.

It is in this latter respect that the rule of law provides the institutional basis for safeguarding both democratic participation and the protection of fundamental rights and freedoms. An independent and impartial judiciary, which ensures citizens due judicial processes and fair and equal treatment under the law, acts as a guarantor of recourse whenever fundamental rights or freedoms could be breached.

# 04 OPPORTUNITIES AND RISKS OF AI AND MACHINE LEARNING AND THEIR IMPACT ON HUMAN RIGHTS, DEMOCRACY, AND THE RULE OF LAW

Artificial intelligence (AI) technologies provide a range of opportunities for the improvement of human lives and the functioning of government. The power, scale, and speed of AI systems can improve efficiency and effectiveness in numerous domains, including healthcare, transport, education, and public administration. They can take over tedious, dangerous, unpleasant, and complicated tasks from human workers. However, AI technologies also have the potential to negatively impact human rights, democracy, and the rule of law. These combined opportunities and risks should be understood in light of AI being "*socio-technical*" – AI is a broad range of sophisticated technologies that operate in human contexts, designed to fulfil human-defined goals. As such, AI technologies can be said to reflect the values and choices of the people who build and use them.

AI can be applied to make predictions about human behaviour, to identify indicators of disease, and to assess risks posed to the interests or well-being of others. All of these tasks may affect the rights, opportunities, and well-being of those upon whom they are used. For this reason, accountability is an essential aspect of developing and using such systems. While AI can take over tedious or complex tasks from humans, the choices involved in the construction and use of AI systems can result in the reproduction of harmful bias and other fallibilities of human judgement that adversely impact affected individuals and wider society in ways that are harder to identify than when done by humans.

So, in addition to evaluating the technical features of a particular system or technology, AI accountability requires that we also thoroughly consider potential harms and benefits for individuals and groups. Among the potential harms is unjust bias, which may occur explicitly, such as when AI models make discriminatory predictions or otherwise treat a particular demographic group or identity differently than others without justification. Assessing AI systems for their potential to cause harm is made more difficult by the opacity of some AI systems. In addition to being constructed using specialised knowledge, the work of AI technologies can be difficult to interpret or explain due to both its technical complexity and intellectual property protections.

The specific human rights implications for AI systems can be viewed through provisions of the European Convention of Human Rights (ECHR) and the European Social Charter (ESC), including its specific guarantees regarding **liberty and justice, privacy, freedom of expression, equality and non-discrimination**, and **social and economic rights**. There are additional implications of AI on democracy and the rule of law that do not fall clearly within the provisions of the ECHR and the ESC but are similarly important considerations nonetheless. A thorough consideration of the risks and opportunities presented by AI systems will help us to identify where existing rights and freedoms provide needed protections, where further clarification of existing rights and freedoms is needed, and where new rights and freedoms must be tailored to the novel challenges and opportunities raised by AI and machine learning.

**Liberty and Justice:** AI can adversely affect the liberty and justice of individuals, particularly when implemented in high impact contexts such as criminal justice. The complexity and opacity of AI systems may interfere with the right to a fair trial including the right to equality of arms, in which a party subject to an algorithmic decision can adequately examine and contest their reasoning. While the use of AI in this context may reduce arbitrariness and discriminatory action, judicial decisions supported or informed by AI may negatively affect the rulemaking and decisional independence of the judiciary. As a result, judicial actors should have a sufficient level of understanding about the AI they use to ensure accountability for decisions made with its assistance.

A system that supports criminal sentencing decisions with scores to represent the risk that a convicted criminal will commit additional crimes must be interpretable, verifiable, and open to challenge by the defendant to ensure a fair and open judicial process.

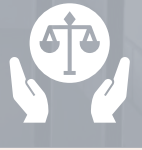

A system that analyses facial expressions, tone of voice, word choice, and other biometric cues and compares them to models to predict whether a job candidate will be a "successful" hire may violate the job candidate's sense of bodily and emotional privacy.

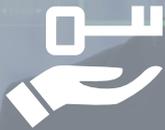

**Privacy:** AI can access enormous amounts of data about individuals and process it with incredible speed. AI can make predictions about a person's behaviour, state of mind, and identity by sensing information that is not necessarily considered personal or private, such as facial expressions, heart rate, physical location, and other seemingly mundane or publicly accessible data. This can have the effect of being invasive of a person's sense of privacy, and can also have so-called "panoptic effects" by causing a person to alter their behaviour upon suspicion it is being observed or analysed.

**Freedom of expression, association, and assembly:** A functioning democracy requires open social and political discourse and the minimisation of undue influence or manipulation by any particular person or institution. AI places these values at risk where it is used to collect and process information about online and offline activity through logging and analysing website and social media usage or extracting information through biometric surveillance. AI used in this way contributes to the sense that one is being watched and listened to, potentially chilling speech and political action. AI use by social media platforms determines what posts and ads are displayed, constructing an experience that exploits individual interests and biases to maintain engagement with the platform while potentially reinforcing divisive, anti-democratic, or violent worldviews. AI is also employed to produce highly realistic but fake videos, fake accounts, and other manufactured content that may impede a person's ability to reach informed opinions based in fact.

Live facial recognition systems may prevent citizens from exercising their freedoms of assembly and association, robbing them of the protection of anonymity and having a chilling effect on social solidarity and democratic participation. AI-enabled biometric surveillance may also strip citizens of their right to informed and explicit consent in the collection of personal data.

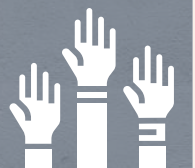

**Equality and Non-Discrimination:** AI systems are capable of reproducing and augmenting the patterns of discriminatory treatment that exist in the society in which they are created and used. This can occur when the stereotyping biases and blind spots of system developers shape the choices made in the design and deployment of systems. It can also occur when historical structures of inequality and discrimination become entrenched in the datasets that are used to train AI and machine learning models. Where AI relies on such biased information, discriminatory human decisions that produced a dataset can lead to discriminatory algorithmic decisions and behaviours.

When predictive policing systems rely on historical data, they risk reproducing the results of prior discriminatory practices. This can lead to "feedback loops", where each new policing decision based on historical data produces new data, leading to members of marginalised groups being disproportionately suspected and arrested.

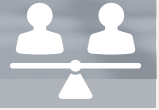

Ride-hailing and delivery services coordinated by mobile apps enable companies to automate the management and supervision of large workforces and to dehumanise labor relations and management practices in turn. This can disempower workers and limit avenues of recourse for employees faced with erroneous or unfair pay or employment decisions issued by algorithmic managers.

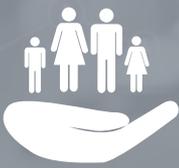

**Social and economic rights:** AI systems are used with increasing frequency by employers and governments in ways that put social and economic rights at risk. Employers use technology to monitor worker behaviour, disrupt unionisation, and to make decisions about hiring, pay, and advancement. In some employment contexts, humans are managed primarily by algorithmic decision systems, potentially affecting their economic opportunities. Likewise, governmental impacts on economic prosperity are implicated where AI is used to allocate public benefits and healthcare. A lack of sufficient oversight of such management may deny benefits to the deserving, threatening their welfare. The automation of both eligibility determination and allocation of government benefits can create more efficient service delivery but can also leave those denied benefits without recourse or leave them to navigate complex forms and other processes without compassionate assistance.

Overlapping with these human rights concerns is the concentration of power that AI affords to its most influential private and public sector developers and implementers. The operators of major online platforms employ AI to choose what content to display and whose voices to make prominent in service of their own, rather than democratic interests. Governments use AI to rank and order information and to monitor and track citizens. Whether done by companies or governments, AI can be used to shape opinions and suppress dissent.

In response to these considerations and concerns, governments should adopt a precautionary approach in the adoption and regulation of AI that balances the realisation of the opportunities presented by AI while ensuring that risks to human beings and human interests are minimised to the extent possible. In contexts where a precautionary approach is found to be insufficient to mitigate risk, governments should consider prohibitions on the use of AI. Where there is uncertainty about the level or impact of potential risks, governments should apply a higher degree of regulatory oversight and monitoring of AI systems and be prepared to prohibit its use.

# 05 PRINCIPLES AND PRIORITIES FOR A LEGAL FRAMEWORK

In September 2019, the Council of Europe's Committee of Ministers adopted the terms of reference for the Ad Hoc Committee on Artificial Intelligence (CAHAI). The CAHAI is charged with examining the feasibility and potential elements of a legal framework for the development, design, and deployment of AI systems, based on Council of Europe standards across the interrelated areas of human rights, democracy, and the rule of law. As a first and necessary step in carrying out this responsibility, the CAHAI's *Feasibility Study*, adopted by its plenary in December 2020, has proposed nine principles and priorities that are intended to underpin such a framework of binding and non-binding legal instruments:

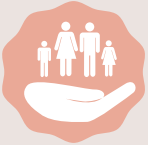
### HUMAN DIGNITY
All individuals are inherently and inviolably worthy of respect by mere virtue of their status as human beings. Humans should be treated as moral subjects, and not as objects to be algorithmically scored or manipulated.

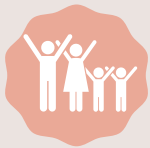
### HUMAN FREEDOM & AUTONOMY
Humans should be empowered to determine in an informed and autonomous manner if, when, and how AI systems are to be used. These systems should not be employed to condition or control humans, but should rather enrich their capabilities.

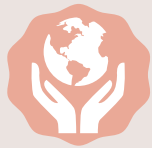
### PREVENTION OF HARM
The physical and mental integrity of humans and the sustainability of the biosphere must be protected, and additional safeguards must be put in place to protect the vulnerable. AI systems must not be permitted to adversely impact human wellbeing or planetary health.

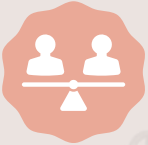
### NON-DISCRIMINATION, GENDER EQUALITY, FAIRNESS & DIVERSITY
All humans possess the right to non-discrimination and the right to equality and equal treatment under the law. AI systems must be designed to be fair, equitable, and inclusive in their beneficial impacts and in the distribution of their risks.

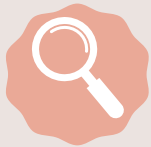
### TRANSPARENCY AND EXPLAINABILITY OF AI SYSTEMS
Where a product or service uses an AI system, this must be made clear to affected individuals. Meaningul information about the rationale underlying its outputs must likewise be provided.

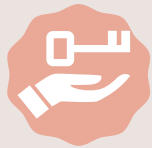
### DATA PROTECTION AND THE RIGHT TO PRIVACY
The design and use of AI systems that rely on the processing of personal data must secure a person's right to respect for private and family life, including the individual's right to control their own data. Informed, freely given, and unambiguous consent must play a role in this.

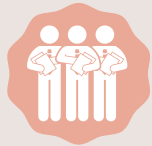
### ACCOUNTABILITY AND RESPONSIBILITY
All persons involved in the design and deployment of AI systems must be held accountable when applicable legal norms are violated or any unjust harm occurs to end-users or to others. Those who are negatively impacted must have access to effective remedy to redress harms.

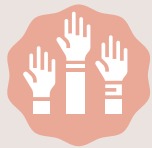
### DEMOCRACY
Transparent and inclusive oversight mechanisms must ensure that the democratic decision-making processes, pluralism, access to information, autonomy, and economic and social rights are safeguarded in the context of the design and use of AI systems.

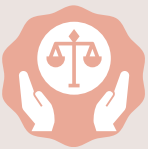
### RULE OF LAW
AI systems must not undermine judicial independence, due process, or impartiality. To ensure this, the transparency, integrity, and fairness of the data, and data processing methods must be secured.

# CONNECTING PRINCIPLES, RIGHTS, AND OBLIGATIONS

These nine principles and priorities are **horizontally applicable**. They apply to the design, development, and deployment of AI systems **across sectors and use cases,** though they could be combined with a sector-specific approach that provides (more detailed) contextual requirements in the form of soft law instruments, such as sectoral standards, guidelines, or assessment lists.

The legal framework is meant to start from this wide-angled point of view. It will aim to secure the nine principles and priorities by identifying **concrete rights** that ensure the realisation of these cross-sectoral principles at the individual level as well as the **key obligations and requirements** that developers and deployers should meet in building and using AI systems that accord with human rights, democracy, and the rule of law. The identified rights could be (1) drawn directly from existing rights, (2) newly established rights that are tailored to the challenges and opportunities raised by AI, or (3) further clarifications of existing rights.

Here is a mapping of how each of the principles and priorities is connected with corresponding rights and obligations:

| | SUBSTANTIVE RIGHTS | KEY OBLIGATIONS |
|---|---|---|
| **HUMAN DIGNITY** | -The right to human dignity, the right to life **(Art. 2 ECHR)**, and the right to physical and mental integrity.<br><br>-The right to be informed of the fact that one is interacting with an AI system rather than with a human being.<br><br>-The right to refuse interaction with an AI system whenever this could adversely impact human dignity. | -Member States should ensure that, where tasks would risk violating human dignity if carried out by machines rather than human beings, these tasks are reserved for humans.<br><br>-Member States should require AI deployers to inform human beings of the fact that they are interacting with an AI system rather than with a human being in any context where confusion could arise. |
| **HUMAN FREEDOM & AUTONOMY** | -The right to liberty and security **(Art. 5 ECHR)**.<br><br>-The right to human autonomy and self-determination. The right not to be subject to a decision based solely on automated processing when this produces legal effects on or similarly significantly affects individuals.<br><br>-The right to effectively contest and challenge decisions informed and/or made by an AI system and to demand that such decision be reviewed by a person.<br><br>-The right to freely decide to be excluded from AI-enabled manipulation, individualised profiling, and predictions. This also applies to cases of non-personal data processing.<br><br>-The right to have the opportunity, when it is not excluded by competing legitimate overriding grounds, to choose to have contact with a human being rather than a robot. | -All AI-enabled manipulation, individualised profiling, and predictions involving the processing of personal data must comply with the obligations set out in the Council of Europe Convention for the Protection of Individuals with regard to Automatic Processing of Personal Data.<br><br>-Member States should effectively implement the modernised version of the Convention ("Convention 108+") to better address AI-related issues.<br><br>-Member States should require AI developers and deployers to establish human oversight mechanisms that safeguard human autonomy, in a manner that is tailored to the specific risks arising from the context in which the AI system is developed and used.<br><br>-Member States should require AI developers and deployers to duly communicate options for redress in a timely manner. |

| SUBSTANTIVE RIGHTS | KEY OBLIGATIONS |
|---|---|
| 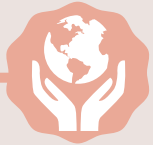**PREVENTION OF HARM**<br><br>-The right to life **(Art. 2 ECHR)** and the right to physical and mental integrity.<br><br>-The right to the protection of the environment.<br><br>-The right to sustainability of the community and biosphere. | -Member States should ensure that developers and deployers of AI systems take adequate measures to minimise any physical or mental harm to individuals, society, and the environment.<br><br>-Member States should ensure the existence of adequate (by design) safety, security, and robustness requirements and compliance therewith by developers and deployers of AI systems.<br><br>-Member States should ensure that AI systems are developed and used in a sustainable manner, with full respect for applicable environmental protection standards. |
| 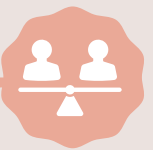**NON-DISCRIMINATION, GENDER EQUALITY, FAIRNESS & DIVERSITY**<br><br>-The right to non-discrimination **(on the basis of the protected grounds set out in Article 14 of the ECHR and Protocol 12 to the ECHR)**, including intersectional discrimination.<br><br>-The right to non-discrimination and the right to equal treatment.<br><br>-AI systems can also give rise to unjust categorisation based on new types of differentiation that are not traditionally protected.<br><br>-This right must be ensured in relation to the entire lifecycle of an AI system (design, development, implementation, and use), as well as to the human choices concerning AI design, adoption, and use, whether used in the public or private sector. | -Member States are obliged to ensure that the AI systems they deploy do not result in unlawful discrimination, harmful stereotypes (including but not limited to gender stereotypes), and wider social inequality, and should therefore apply the highest level of scrutiny when using or promoting the use of AI systems in sensitive public policy areas, including but not limited to law enforcement, justice, asylum and migration, health, social security, and employment.<br><br>-Member States should include non-discrimination and promotion of equality requirements in public procurement processes for AI systems and ensure that the systems are independently audited for discriminatory effects prior to deployment.<br><br>-Member States should impose requirements to effectively counter the potential discriminatory effects of AI systems deployed by both the public and private sectors and protect individuals from the negative consequences thereof. Such requirements should be proportionate to the risks involved.<br><br>-Member States should encourage diversity and gender balance in the AI workforce and periodic feedback from a diverse range of stakeholders. Awareness of the risk of discrimination, including new types of differentiation, and bias in the context of AI should be fostered. |

| SUBSTANTIVE RIGHTS | KEY OBLIGATIONS |
|---|---|
| 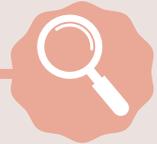**TRANSPARENCY & EXPLAINABILITY**<br><br>-The right to be promptly informed that a decision which produces legal effects or similarly significantly impacts an individual's life is informed or made by an AI system **(Convention 108+)**.<br><br>-The right to a meaningful explanation of how such an AI system functions, what optimisation logic it follows, what type of data it uses, and how it affects one's interests, whenever it generates legal effects or similarly impacts individuals' lives. The explanation must be tailored to the context and provided in a manner that is useful and comprehensible for an individual, allowing individuals to effectively protect their rights.<br><br>-The right of a user of an AI system to be assisted by a human being when an AI system is used to interact with individuals, in particular in the context of public services. | -Users should be clearly informed of their right to be assisted by a human being whenever using an AI system that can impact their rights or similarly significantly affect them, particularly in the context of public services, and of how to request such assistance. Member States should require developers and deployers of AI systems to provide adequate communication.<br><br>-Whenever the use of AI systems risks negatively affecting human rights, democracy, or the rule of law, Member States should impose requirements on AI developers and deployers regarding traceability and the provision of information.<br><br>-Member States should make public and accessible all relevant information on AI systems (including their functioning, optimisation functioning, underlying logic, type of data used) that are used in the provision of public services, while safeguarding legitimate interests such as public security or intellectual property rights, yet securing the full respect of human rights. |
| 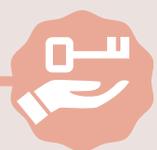**DATA PROTECTION & RIGHT TO PRIVACY**<br><br>-The right to respect for private and family life and the protection of personal data **(Art. 8 ECHR)**.<br><br>-The right to physical, psychological, and moral integrity in light of AI-based profiling and emotion/personality recognition.<br><br>-All the rights enshrined in Convention 108+ and in its modernised version, and in particular with regard to AI-based profiling and location tracking. | -Member States must ensure that the right to privacy and data protection are safeguarded throughout the entire lifecycle of AI systems that they deploy, or that are deployed by private actors.<br><br>-Member States should take measures to effectively protect individuals from AI-driven mass surveillance, for instance through remote biometric recognition technology or other AI-enabled tracking technology.<br><br>-When procuring or implementing AI systems, Member States should assess and mitigate any negative impact on the right to privacy and data protection as well as on the broader right to respect for private and family life. Of particular concern is the proportionality of the system's invasiveness in light of the legitimate aim it should fulfil, as well as its necessity to achieve it.<br><br>-Member states should put in place appropriate safeguards for transborder data flows to ensure that data protection rules are not circumvented. |

| SUBSTANTIVE RIGHTS | KEY OBLIGATIONS |
|---|---|
| -The right to an effective remedy for violation of rights and freedoms **(Art. 13 ECHR)**.<br><br>-This should also include the right to effective and accessible remedies whenever the development or use of AI systems by private or public entities causes unjust harm or breaches an individual's legally protected rights. | -Member States must ensure that effective remedies are available under respective national jurisdictions, including for civil and criminal responsibility, and that accessible redress mechanisms are put in place for individuals whose rights are negatively impacted by the development or use of AI applications.<br><br>-Member States should establish public oversight mechanisms for AI systems that may breach legal norms in the sphere of human rights, democracy, or the rule of law.<br><br>-Member States should ensure that developers and deployers of AI systems (1) identify, document, and report on potential negative impacts of AI systems on human rights, democracy, and the rule of law; and (2) put in place adequate mitigation measures to ensure responsibility and accountability for any harm caused.<br><br>-Member States should put in place measures to ensure that public authorities are always able to audit AI systems used by private actors, so as to assess their compliance with existing legislation and to hold private actors accountable. |

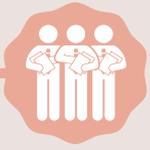

**ACCOUNTABILITY & RESPONSIBILITY**

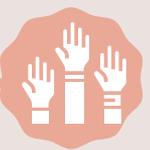

**DEMOCRACY**

| | |
|---|---|
| -The right to freedom of expression, freedom of assembly and association **(Art. 10 and 11 ECHR)**.<br><br>-The right to vote and to be elected, the right to free and fair elections, and in particular universal, equal and free suffrage, including equality of opportunities and the freedom of voters to form an opinion. In this regard, individuals should not to be subjected to any deception or manipulation.<br><br>-The right to (diverse) information, free discourse, and access to plurality of ideas and perspectives.<br><br>-The right to good governance. | -Member States should take adequate measures to counter the use or misuse of AI systems for unlawful interference in electoral processes, for personalised political targeting without adequate transparency, responsibility, and accountability mechanisms, or more generally for shaping voters' political behaviours or to manipulate public opinion.<br><br>-Member States should adopt strategies and put in place measures for fighting disinformation and identifying online hate speech to ensure fair informational plurality.<br><br>-Member States should subject their public procurement processes to legally binding requirements that ensure the responsible use of AI in the public sector by safeguarding compliance with the above-mentioned principles, including transparency, fairness, responsibility, and accountability.<br><br>-Member States should put in place measures to increase digital literacy and skills in all segments of the population. Their educational curricula should adjust to promote a culture of responsible innovations that respects human rights, democracy, and the rule of law. |

| SUBSTANTIVE RIGHTS | KEY OBLIGATIONS |
|---|---|
| -The right to a fair trial and due process **(Art. 6 ECHR)**. This should also include the possibility of receiving insight into and challenging AI-informed decisions in the context of law enforcement or justice, including the right to review of such decision by a human.<br><br>-The right to judicial independence and impartiality, and the right to legal assistance.<br><br>-The right to an effective remedy **(Art. 13 ECHR)**, also in case of unlawful harm or breach an individual's human rights in the context of AI systems. | -Member States must ensure that AI systems used in the field of justice and law enforcement are in line with the essential requirements of the right to a fair trial. To this end, they should ensure the quality and security of judicial decisions and data, as well as the transparency, impartiality, and fairness of data processing methods. Safeguards for the accessibility and explainability of data processing methods, including the possibility of external audits, should be introduced to this end.<br><br>-Member States must ensure that effective remedies are available and that accessible redress mechanisms are put in place for individuals whose rights are violated through the development or use of AI systems in contexts relevant to the rule of law.<br><br>-Member States should provide meaningful information to individuals on the use of AI systems in the public sector whenever this can significantly impact individuals' lives. Such information must especially be provided when AI systems are used in the field of justice and law enforcement, both as concerns the role of AI systems within the process, and the right to challenge the decisions informed or made thereby.<br><br>-Member States should ensure that use of AI systems does not interfere with the decision-making power of judges or judicial independence and that any judicial decision is subject to meaningful human oversight. |

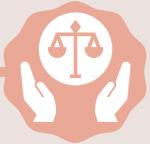

**RULE OF LAW**

# ADDITIONAL CONSIDERATIONS FOR PRINCIPLES, RIGHTS, AND OBLIGATIONS

There are some additional factors that should be weighed when the potential introduction of new rights and obligations in a future principles-based legal framework on AI systems is being considered. First, these rights and obligations should be necessary, useful, and proportionate to the goal of protecting citizens from the negative impacts of AI systems on human rights, democracy, and the rule of law, while at the same time ensuring the just and equitable distribution of their benefits. These considerations of risks and benefits should be comprehensive and should incorporate an awareness of the balance of legitimate interests at stake. A risk-based and benefits-aware approach should also differentiate between different levels of risk and take this into account when regulatory measures are formulated and agreed.

### Main Elements of a Risk-Based and Benefits-Aware Approach:

- Consider **use context** and the **potential impact** of the AI technology
- Consider **domain of application** and **affected stakeholders**
- **Assess and review risks regularly** and **systematically, tailoring** any **mitigating measures** to these risks
- **Optimise societal benefits** of AI innovation by **targeting regulatory measures** in this risk-based way

In terms of obligations and requirements, national authorities should play a central role in systematically assessing domestic legislation to verify its compliance with the principles and priorities of aligning AI design and use with human rights, democracy, and the rule of law, and to identify any legal gaps. Moreover, national mechanisms for the audit and oversight of AI systems should safeguard against harmful instances of non-compliance. Finally, as private actors are increasingly providing critical digital infrastructure for the public sector that affects the public interest, they have a responsibility to align the design, development, and deployment of their technologies with these principles and priorities.

# 06 LANDSCAPE OF LEGAL INSTRUMENTS

## INTERNATIONAL LEGAL FRAMEWORKS

Currently, there are no international laws which focus specifically on AI – or automated decision-making – but a number of existing legal frameworks are relevant. In particular (as summarised above):

- The European Convention of Human Rights (ECHR)
- The European Social Charter (ESC)
- The International Bill of Human Rights
- The Charter of Fundamental Rights of the European Union (CFR)

These legal instruments set out people's fundamental rights, many of which are relevant to applications of AI, for example: The right to non-discrimination and the right to privacy.

Similarly, there are a number of legal instruments which identify people's rights in relation to particular sectors and/or activities, including cybercrime, biomedicine, and aviation. As AI is increasingly used across diverse sectors and in ways which affect more and more parts of our lives, it is increasingly relevant to each of these areas of law.

AI is also relevant to legal instruments which serve to protect vulnerable or minority groups. As such, while there is no specific legal mechanism relating to AI, an increasing number of current legal mechanisms are relevant to the ways in which it is developed and deployed.

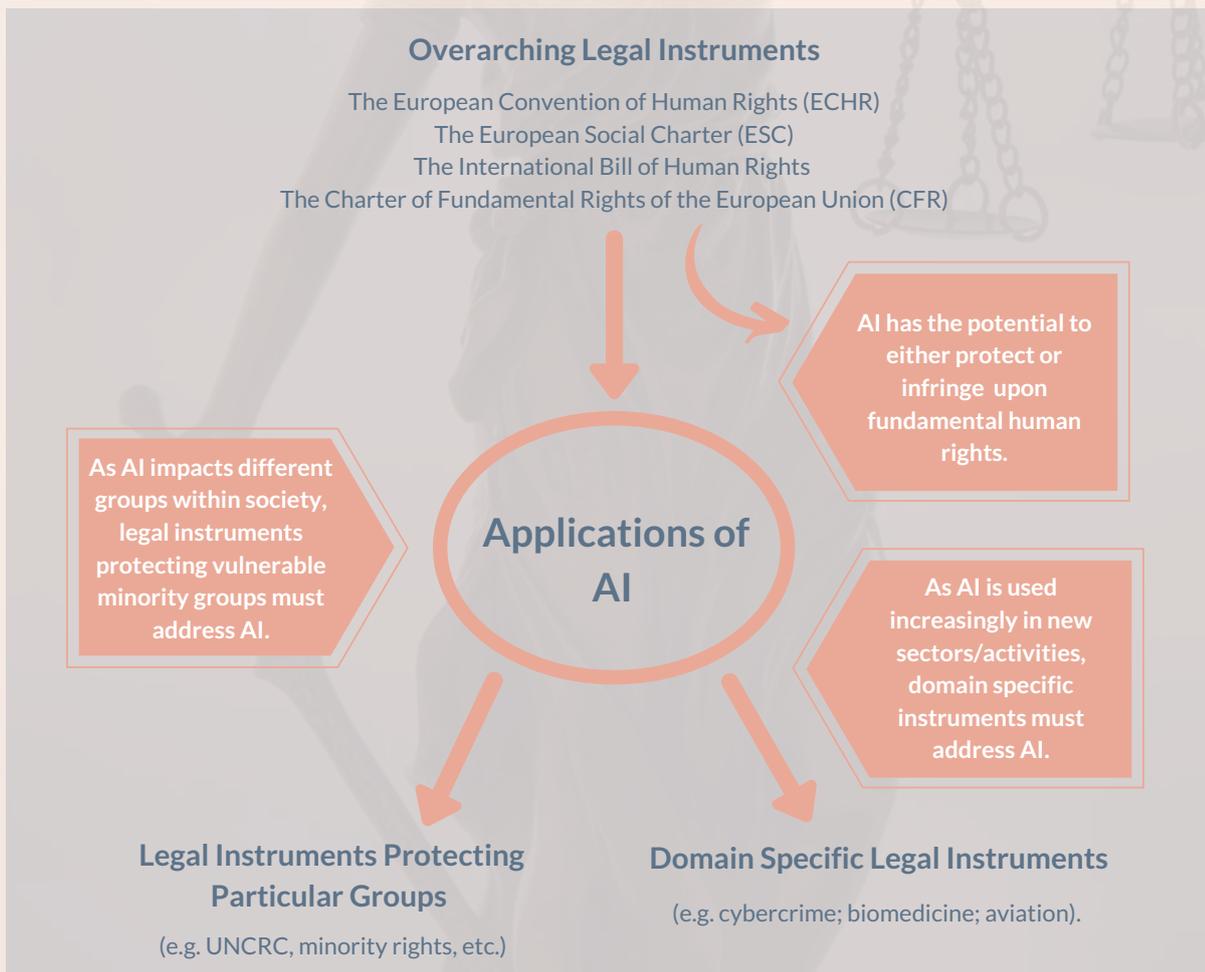

# Current Soft Law Approaches

Currently the main approaches to governance or regulation of AI reflect "soft law" approaches. The difference between hard and soft law can be viewed below.

Recent years have brought a proliferation of sets of guidance and principles for ethical practice relating to AI. These are typically aimed at demonstrating trustworthiness in the ways that AI is developed and deployed. Such guidance or principles have been developed by private sector, academic, and public sector organisations. In many cases, the development of internal guidance and best practice has served as a means of arguing against the need for hard law relating to AI or greater centralised regulation of AI. Many organisations who have proposed principles or guidance for ethical AI have argued strongly in favour of self-regulation.

Voluntary codes of conduct adopted within organisations using AI can play an important role in shaping organisational culture and lead to meaningful impacts on practice. Moreover, they have advantages in their flexibility, adaptability, immediacy of implementation, broader appeal, and capacity to be reviewed and amended quickly. However, they have also been criticised for being tokenistic and largely rhetorical.

There is some consistency in the principles put forward within existing sets of guidance. For example, transparency is routinely emphasised. By contrast there is a lack of consistency around practical guidance. This leads to very different approaches being taken and varied understandings of what is ethically required or how AI should be regulated. Additionally, while there is no shortage of codes of practice or guidance around ethical AI, there is generally a lack of accountability and transparency relating to enforcement of these. Enforcement via internal committees or review panels has been criticised for lacking transparency or effectiveness.

As such there is a strong case for combining voluntary soft law approaches with mandatory governance.

### Hard law
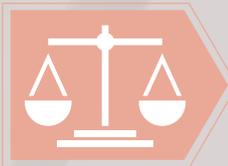
- Legally binding instruments
- Fixed sanctions
- Enforceable through litigation and court proceedings

### Soft law
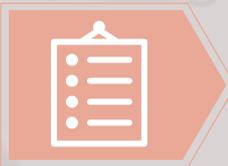
- Non-binding recommendations, guidelines, certifications, or declarations that consolidate common principles and standards of best practices
- Often open to interpretation
- No legal sanctions

# National Legal Instruments

Internationally there is growing interest in developing approaches to govern or regulate AI. Soft law approaches dominate. A consultation with CAHAI members found that:
- 30 member states and 4 observer states have strategies and policies relating to AI systems;
- 1 member state has launched a voluntary AI certification programme;
- 2 member states have formally endorsed international or European non-binding AI ethics frameworks;
- 12 member states and 4 observer states have adopted one or more instruments.

These approaches have been led by a variety of institutions including national councils, committees, specialist AI public institutions, and government entities.

In terms of developing hard law, the consultation with CAHAI members found that:
- 4 member states have adopted specific legal frameworks on AI in the testing and use of autonomous vehicles (self-driving cars);
- 2 member states are developing legal frameworks on the use of AI in recruitment and automated decision-making by public authorities.

## THE ROLE OF PRIVATE ACTORS

Private actors (e.g. businesses) have significantly shaped the field of AI ethics, including through the creation and adoption of voluntary codes of conduct. In some cases private actors have also argued in favour of a regulatory framework to enhance legal certainty around AI.

It is clear that private actors have an important role to play. Private actors' responsibility to respect human rights across their operations, products, and services is set out in the U.N. Guiding Principles on Business and Human Rights.

If a new regulatory approach is implemented, the involvement and cooperation of private actors will be vital to develop sectorial soft law. This will be important to complement and support the implementation of hard law in context-specific manners (for example through sector-specific guidance or certification schemes).

An effective regulatory framework for AI will require close cooperation between all stakeholders, including states, public sector bodies, civil society, and business in order to reflect diverse interests and perspectives.

## CURRENT LIMITATIONS

Many of the legal instruments currently used to regulate aspects of AI were developed before AI systems became commonplace. As such they may be inadequate to deal with the various impacts and risks of AI.

Soft law approaches are non-binding and rely on voluntary compliance which can lead to varied practices and outcomes. Additionally, the varied approaches taken by organisations following soft law can lead to tokenistic or cosmetic commitments to ethical AI. Nonetheless, much work now being done in the area of standards and certification may support future statutory interventions.

There are additionally some important principles which are not currently legally assured in the governance of AI. For example, the need to ensure sufficient human control and oversight, and the effective transparency and explainability of AI systems. There is a lack of legal instruments to address these important technologically specific factors of AI.

While current legal mechanisms, to some extent, protect individual rights, the societal dimensions of AI's risks are not yet sufficiently addressed (e.g., risks to electoral processes or democratic institutions). Protecting democracy and the rule of law requires public oversight and involvement in the responsible design, development, and use of AI systems.

Finally, current regulatory gaps create uncertainty and ambiguity around AI. This is important for AI developers, implementers, and users as well as wider society. Uncertainty in this area is liable to hamper the benefits of AI innovation and may stand in the way of the important innovation which could otherwise benefit citizens and the communities in which they live.

# FUTURE NEEDS AND OPPORTUNITIES

Future regulatory approaches should address the limitations set out above. They should cut across sectors and contain binding provisions to safeguard human rights, democracy, and the rule of law, and to ensure more comprehensive protection. This could complement existing sector-specific rules.

Developing a legally-binding instrument based on Council of Europe standards – should this option be supported by the Committee of Ministers – would contribute to making the Council of Europe initiative unique among other international initiatives, which either focus on elaborating a different type of instrument or have a different scope or background.

# OPTIONS FOR A LEGAL FRAMEWORK

There are several ways in which the Council of Europe could decide to create rules for AI in order to protect human rights, democracy, and the rule of law. Each approach has benefits and drawbacks in terms of expected outcomes.

There are two main distinctions to consider. The first is between binding and non-binding legal instruments, which concerns whether States are bound to the rules that the Council decides upon. The second is how much to consolidate and modernise existing instruments and how much to create entirely new ones. See the graphic below for a map of these approaches and where to find more information in this section.

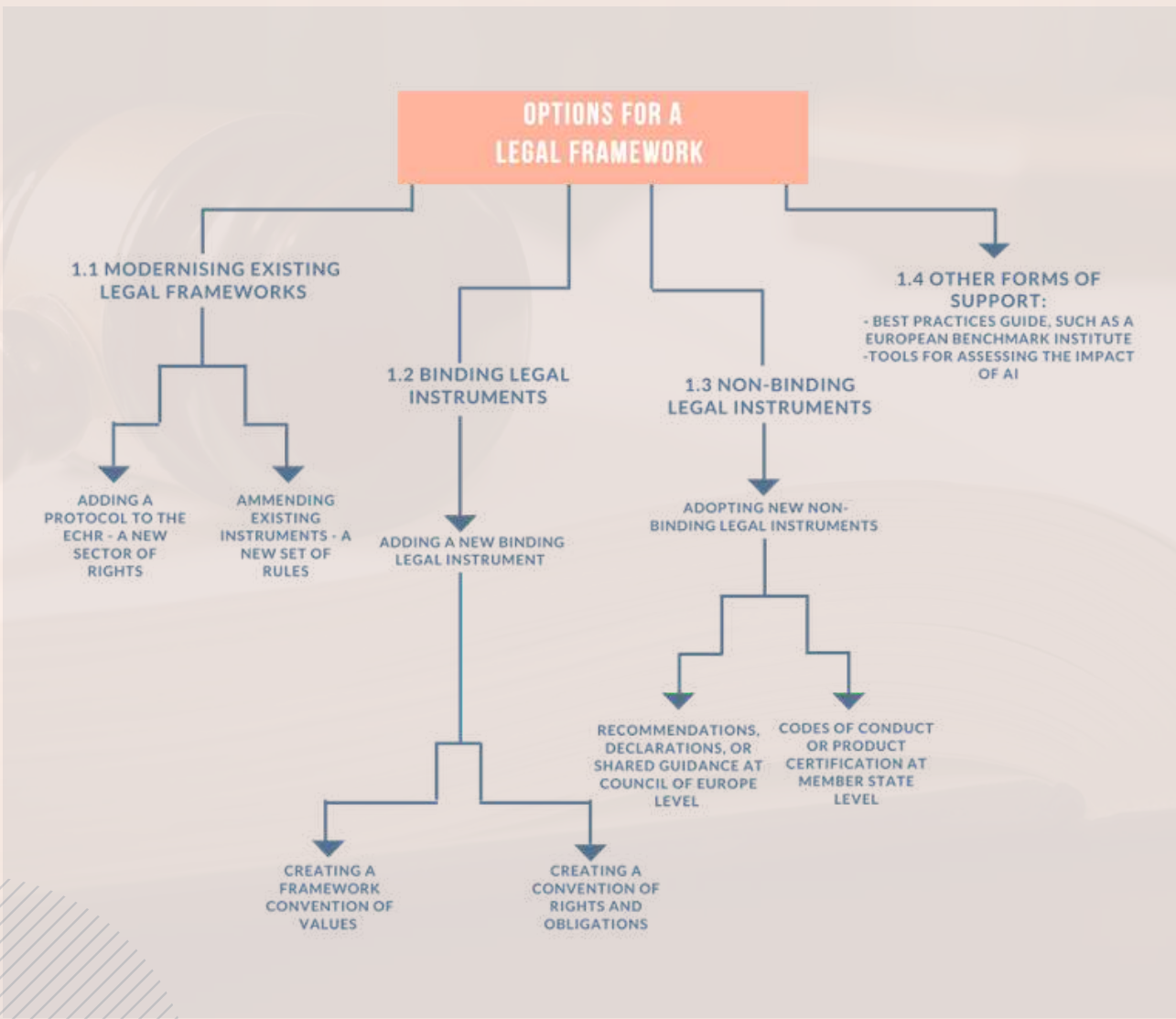

*1.1: Modernising existing binding legal instruments*

One option under consideration is to amend existing rules for the context of AI. For example, this could involve adding a *protocol* (a set of rights) to the existing European Convention on Human Rights. An additional protocol would be a strong statement by Member States of support for the protection of human rights, democracy, and the rule of law in the case of AI, but by itself, would not allow more specific requirements or standards to be laid out. Additional protocols are only binding on States that ratify them, which may make oversight more fragmented. The European Court of Human Rights is, moreover, already over-burdened with cases.

Alternatively, the Council could decide to amend existing *instruments* (sets of rules) to encompass the considerations raised by AI. Two existing instruments that could be amended in this way are the Budapest Convention on Cybercrime, and Convention 108+, which safeguards the processing of personal data about individuals. An advantage of this approach is that there is existing capacity for monitoring and enforcing the rules that are already in place. However, one drawback of this approach is that it would be difficult to adapt the existing instruments sufficiently. The challenges of cybercrime and data protection are related, but not identical, to those raised by AI, such as accountability and explainability for automated systems.

A final consideration is that these two options could be *combined* to address the drawbacks of each. Adding a protocol could establish overall principles and values, and amending existing instruments could provide more detail about the obligations of States to protect these principles in practice, while ensuring there is sufficient capacity for overseeing this. The question is whether a combined approach would be too slow and unwieldy, set against the fast pace of AI development and adoption.

*1.2: Adopting a new binding legal instrument*

An alternative approach would be to develop and adopt an entirely new set of binding rules specifically for AI. There are two forms this could take, a *convention* or a *framework convention*. Similar to the distinction between *protocols* and *instruments* above, a *framework convention* sets out broad principles and areas for action, whereas a *convention* regulates a specific matter in a concrete way through the creation of rights and obligations. However, as treaties they have the same status in terms of international law. Let's look at each in turn.

A *framework convention* could provide broad principles and core values to be respected in the design and rollout of AI systems, but it would leave significant discretion to States as to how these principles and values would be implemented in practice. After the framework convention was established, signers to the convention could decide to create more detailed protocols and specific provisions. This approach could be well-suited to the rapid development of AI and the novel ethical issues that it poses. A framework convention could include agreed upon principles and rules for AI development, as well as specific guidance about how to ensure oversight and cooperation between countries. Similar agreements are already in place among Council of Europe Members for protecting national minorities, and protecting people in the context of medical treatment and experimentation — which is notable because both issues have some overlap with the potential harms of AI systems. Typically, however, framework conventions only identify general duties for States rather than concrete rights for people, giving States leeway in how the principles are implemented.

*Conventions* can allow for more comprehensive regulation. In the case of AI, a convention could identify the rights and obligations that would safeguard human rights, democracy, and the rule of law, and give greater legal protection to people as a result. Taking the convention route would encourage States to act urgently to introduce relevant national laws, and it would create a level playing field for responsible, trustworthy AI products, even across national borders.

The risk with taking the convention route, however, is that it could be overly rigid and impair novel uses of AI that may benefit society. Nonetheless, a concrete set of internationally binding rules would provide legal certainty to all involved, provide strong protection for individuals adversely affected by AI, and lay the foundations for truly responsible AI development.

Regardless of whether a *framework convention* or *convention* is chosen, the addressees of this instrument (that is, those who the rules are chiefly aimed at) would be States, who by formally adopting the convention would agree to become bound by their terms under international law. However, the timeline for getting a convention adopted is unclear, and even States who voted in favour of it at the Council of Europe would not be obliged to formally adopt it. Additionally, it would be important to ensure that other actors such as nations outside Europe adopt equivalent rules, otherwise international rules and standards for AI may become fragmented.

1.3 Non-binding legal instruments

Non-binding or "soft law" instruments do not have the force of international law behind them but may nonetheless play a role guiding States and other actors in a positive direction. Although soft law cannot by itself ensure that AI is oriented towards human rights, democracy, and the rule of law, it can contribute to this effort, and has the advantages of being flexible, adaptable, and quick to implement. Non-binding legal instruments can be divided into those that are enacted at the level of the Council of Europe and those to be approved by Member States. These aren't mutually exclusive, but again, let's look at each in turn.

A broad soft law instrument at the *Council of Europe* level could take the form of a recommendation or a declaration, either as a stand-alone document or to complement one of the binding instruments discussed above. Another option is to create guidance documents or manuals that help shed light on the implications of AI for human rights, democracy, and the rule of law. These documents would be developed with all relevant parties, including representatives of government, the private sector, civil society, and academia, and would be "evolving", updated over time to reflect new developments.

At the M*ember State level*, soft law instruments could take the form of guidelines, codes of conduct, or labels, marks, or seals of certification for AI products. These examples of soft law could be incorporated into the governance, procurement, and auditing practices of organisations such as private companies. However, while this form of "self-regulation" could complement other principles and rules, it should not stand in for or replace the obligations of Member States to actively safeguard human rights, democracy, and the rule of law.

1.4 Other forms of support

Beyond binding and non-binding legal instruments, other forms of support could be provided to Member States and other actors. This includes the potential for best practices to be established to help guide positive action. Creating a "European Benchmarking Institute" could be an effective way to identify and build consensus around what these best practices should be and how they should be supported. In addition, creating a model or tool that allows for assessing the impact of AI at the Council of Europe level could help to bring the implementation of standards and values about AI across the continent to the same level.

To summarise, any approach to effectively ensuring that AI safeguards democracy, human rights, and the rule of law is likely to require a combination of the horizontal (binding and non-binding) approaches outlined here and more sector-specific principles, standards, and requirements.

# 07 PRACTICAL MECHANISMS TO SUPPORT THE LEGAL FRAMEWORK

What practical mechanisms are available to help support the effectiveness of the legal framework, ensure compliance, and promote best practices? We'll now explore some answers to this question by looking at the role of the mechanisms and the relevant actors, and then outlining some examples of mechanisms to a) support compliance and b) to support follow-up activities.

## THE ROLE OF COMPLIANCE MECHANISMS

There are a variety of practical mechanisms that are designed to support and ensure compliance, including human rights due diligence, impact assessments, certification and standards, auditing and monitoring, and even regulatory sandboxes. These mechanisms support compliance with the legal framework, but also confer additional benefits such as increased transparency and trust. They also promote best practices within and across industries, such as the reflective and anticipatory assessment of an AI-enabled system, from the earliest stages of project design to ongoing mechanisms for monitoring the system following its deployment.

The legal framework should set high-level requirements for how to develop these mechanisms. For instance, it may suggest that the use of compliance mechanisms should evolve, alongside the development and deployment of a system, to account for any changes in its function.

While the legal framework should set principles-based requirements for how to develop compliance mechanisms, it should remain the responsibility of Member States to implement them based on existing roles of local institutions and regulatory culture.

### From Compliance to Assurance

Practical mechanisms can also be used to provide *assurance* to relevant operators or users, as well as to promote best practices. This framing extends the role of practical mechanisms beyond a *mere compliance* perspective, and helps to promote an assurance ecosystem that has myriad benefits including:

- assisting internal **reflection** and **deliberation** by providing practical means for evaluating the design, development, and deployment of AI-enabled systems or products, using a dynamic approach that evolves alongside the system (e.g. monitoring changes in the behaviour of the system post-deployment)
- facilitating **transparent communication** between developers, assurers, operators and users, and wider stakeholders
- supporting processes of **documentation** (or reporting) to **ensure accountability** (e.g. audits)
- building **trust and confidence** by promoting and adopting best practices (e.g. standards or certification schemes).

## THE ROLE OF DIFFERENT ACTORS

At a broad level, the following three categories help identify actors that can each contribute, in a complementary way, to ensuring national regulatory compliance.

| ACTOR | ROLE OF ACTOR |
|---|---|
| Assurers of Systems | Independent oversight bodies, such as expert committees, sectoral regulators, or private sector auditors should represent and be accountable to clearly identified stakeholder groups affected by practical applications of AI. However, their scope should not be expected to cover all AI-based products and systems. |
| Developers of Systems | Private and public sector developers can support compliance by adopting policies that increase the visibility of where such technologies are being deployed (e.g. by publishing public sector contracts, or by establishing public registers or notification systems). Standardised tools for internal audit and self-certification have limitations but can also help. |
| Operators and Users of Systems | Well informed operators and users of AI generate demand and can use this purchasing power to incentivise AI application providers and vendors to comply with the future legal framework. This is particularly true of the public sector and its significant procurement power. |

It should also be noted that many AI systems, and the data flows they rely on, are deployed across multiple jurisdictions making it necessary to ensure that adequate mechanisms for information sharing and reporting are in place to support the tasks of the relevant actors.

## EXAMPLES OF TYPES OF COMPLIANCE MECHANISMS

There are a wide variety of compliance mechanisms. Some will work best in certain contexts (e.g. different regulatory cultures) and depending on the various components of an AI system that are subject to compliance (e.g. features of the training data). To help determine the mechanisms that are best suited to each context, inclusive and participatory processes should be carried out with the relevant stakeholders.

There are some shared characteristics of effective practical mechanisms, which a legal framework could specify as principles that should be adhered to. These could include:

- **Dynamic (not static) assessment** at the start and throughout the AI project lifecycle to account for ongoing decision-making
- Mechanisms should be **technology adaptive** to support efforts at future-proofing
- The processes and outputs of the mechanisms should be **differentially accessible** and **understandable** to experts and non-experts to support appeals and redress
- There should be **independent** oversight by the appropriate body or party (e.g. auditor)
- **Evidence-based** technical standards, certifications, and practices should be promoted and used

The following set of mechanisms represents a toolkit that meets many of these principles, while also providing opportunity for refinement and regulatory innovation.

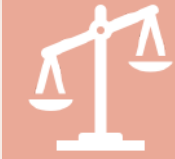
### Human rights due diligence
To ensure that the design, development, and deployment of AI systems do not violate human rights, it is vital that organisations exercise due diligence. The use of impact assessments is one practical means for identifying, preventing, mitigating, and accounting for adverse human rights impacts that may arise from the use of AI-enabled systems. The effective use of impact assessments will depend on the socioeconomic indicators used and the data that are collected. For instance, an impact assessment may want to explore the impact that an AI-enabled system has on individual well-being, public health, freedom, accessibility of information, socioeconomic inequality, environmental sustainability, and more.

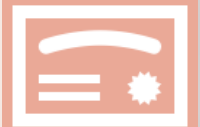
### Certification and quality labelling
Standards and certification schemes are widely used as indicators of safety and quality and could be extended to AI-enabled systems (e.g. certifying that a particular system has undergone extensive evaluation and testing, based on industry standards). The scope of such schemes could apply to either the products and systems themselves or to the organisations responsible for developing the products or systems.

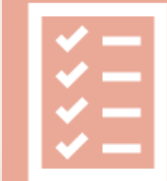
### Auditing
Regular audits by independent, expert bodies with responsibilities for overseeing a particular industry (e.g. healthcare) or domain (e.g. autonomous vehicles) can help facilitate a move towards more transparent and accountable use of AI-enabled systems.

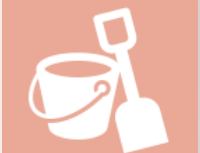
### Regulatory sandboxes
The use of regulatory sandboxes enables authorised firms the opportunity to test AI-enabled products or systems, which are not protected by current regulation, in a safe and controlled manner (i.e. within a sandbox). The use of regulatory sandboxes can help reduce the time-to-market and lower costs for the organisation, supporting innovation in a controlled manner.

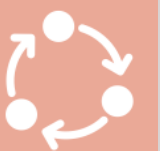
### Continuous, automated monitoring
Once deployed, the behaviour of AI systems needs continuous monitoring to ensure that the functionality of the system continues as expected. There are means by which the process of monitoring can be automated to ensure that any drift in the functionality of an AI-enabled system is identified and addressed as early as possible. However, the use of automated monitoring also carries risk due to the potential loss of human oversight or potential for deskilling of professional compliance checkers.

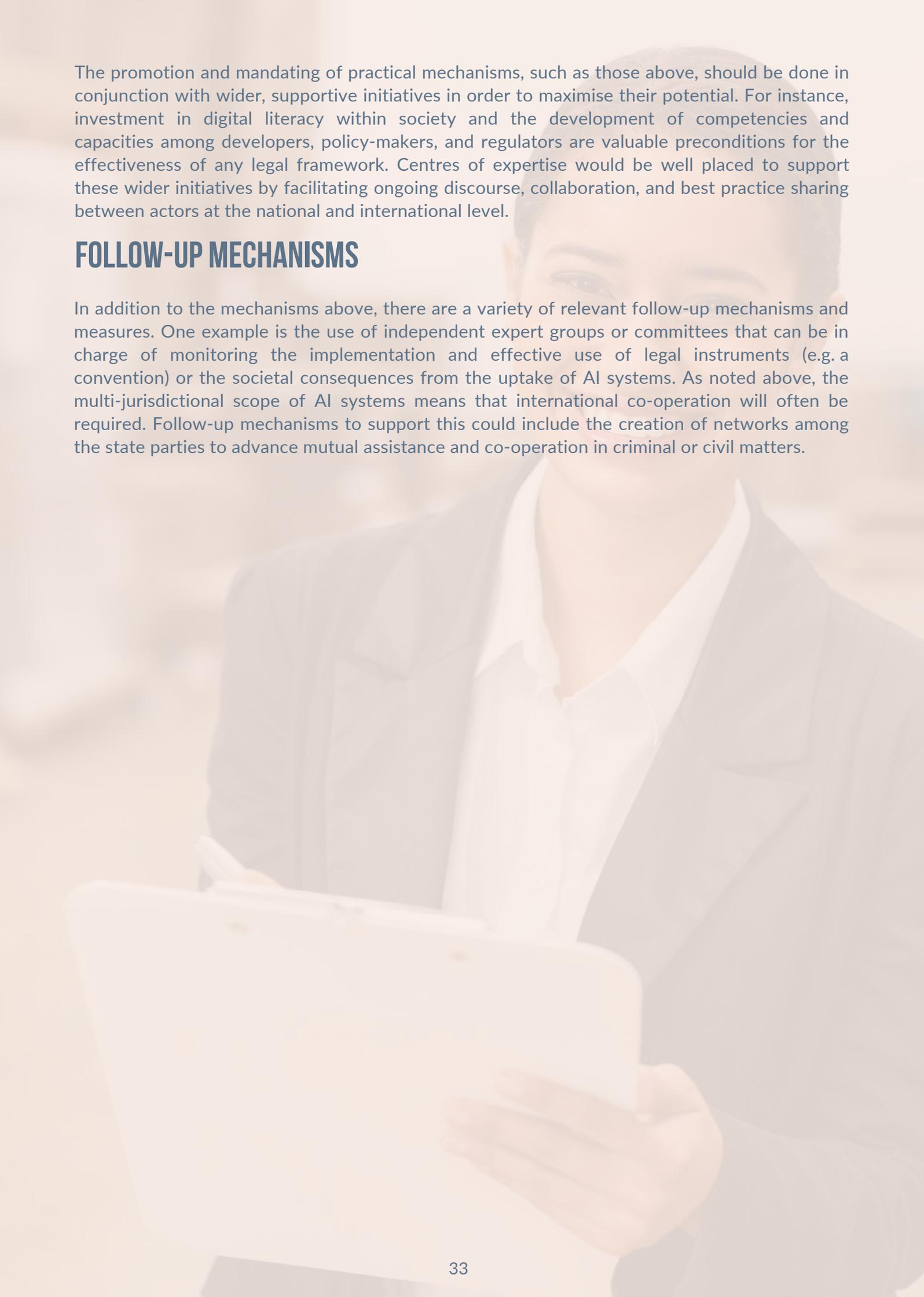

The promotion and mandating of practical mechanisms, such as those above, should be done in conjunction with wider, supportive initiatives in order to maximise their potential. For instance, investment in digital literacy within society and the development of competencies and capacities among developers, policy-makers, and regulators are valuable preconditions for the effectiveness of any legal framework. Centres of expertise would be well placed to support these wider initiatives by facilitating ongoing discourse, collaboration, and best practice sharing between actors at the national and international level.

## FOLLOW-UP MECHANISMS

In addition to the mechanisms above, there are a variety of relevant follow-up mechanisms and measures. One example is the use of independent expert groups or committees that can be in charge of monitoring the implementation and effective use of legal instruments (e.g. a convention) or the societal consequences from the uptake of AI systems. As noted above, the multi-jurisdictional scope of AI systems means that international co-operation will often be required. Follow-up mechanisms to support this could include the creation of networks among the state parties to advance mutual assistance and co-operation in criminal or civil matters.

# 08 CONCLUSION

In this primer, we have tried to introduce the main elements of the CAHAI's *Feasibility Study*, and we have provided some background information about the technical aspects of AI and the interwoven relationship of human rights, democracy, and the rule of law. We hope that, taken together, this material can function as a kind of launching pad for meaningful reflection on the prospects for a principles-based legal framework for governing AI research and innovation in accordance with the Council of Europe's stewardship of fundamental rights and freedoms, justice, and democratic values. Setting these transformative and increasingly powerful technologies on the right path for both citizens and wider society will demand well-informed, visionary policy-making and diligent anticipatory reflection. The *Feasibility Study*, and this supporting primer, offer first steps in this direction.

As the work of the CAHAI now enters the stakeholder consultation and outreach phase, it must be emphasised that the quality and success of this important effort will now depend on the wisdom and insights of as wide and inclusive a group of participants as possible. This reliance on you, the reader, at this critical stage makes good sense. The democratic steering of technology, and technology policy, is at the very heart of the human centred and values-driven perspective that places human rights, democracy, and the rule of law in the pole position for shaping the future of AI governance and digital innovation, more generally. It is, in fact, only through ample feedback and critique, that the voices of impacted individuals and communities can be properly heard and heeded. It is through scrupulous stakeholder consultation alone that lived experience can properly inform this cooperative endeavour to ensure the development of a sustainable technological ecosystem that safeguards the flourishing of the society of tomorrow.

# 09 APPENDICES

## APPENDIX 1: GLOSSARY

- **Accountability:** Accountability can be broken down into two subcomponents: answerability and auditability. Answerability refers to establishing a continuous chain of human responsibility across the whole AI project delivery workflow and demands that explanations and justifications of both the content of algorithmically supported decisions and the processes behind their production be offered by competent human authorities in plain, understandable, and coherent language. Auditability answers the question of how the designers and implementers of AI systems are to be held accountable. This aspect of accountability has to do with demonstrating both the responsibility of design and use practices and the justifiability of outcomes.

- **Algorithm:** An algorithm is a procedure or series of steps that provide instructions on how to take a series of inputs and produce an output. For instance, a recipe can be thought of as an algorithm that provides instructions for taking a series of inputs (i.e. the ingredients) and creating an output (e.g. a cake). In the case of machine learning, the algorithm is typically a series of instructions that instruct a software package to take a dataset (i.e. the input) and learn a model or discover some underlying pattern (i.e. the output).

- **Algorithmic Audits:** There are a variety of approaches to algorithmic auditing, which range from the targeted assessment of a system according to some metric (e.g. level of bias) to a broader approach that focuses on whether the system complies with a set of norms or regulatory area. While typically performed by professionals for the purpose of independent assessment, algorithmic audits have also been used by journalists, academics, and activists as a means of securing greater levels of transparency and accountability.

- **Automated decision:** An automated decision is the selection of an action or a recommendation made using computational processes. Automated decisions describe those that either augment or replace decisional work typically performed by humans alone. Most commonly, automated decisions are *predictions* about persons or conditions in the world derived from machine learning analysis of data about past events and its similarity to a given set of conditions.

- **Automated decision system:** An automated decision system (ADS) augments or replaces human decision-making by using computational processes to produce answers to questions either as discrete classifications (e.g. yes, no; male, female, non-binary; malignant, benign) or continuous scores (e.g. degree of creditworthiness, risk of crime occurrence, projected tumour growth). Most ADS produce *predictions* about persons or conditions using machine learning and other computational logic by calculating the probability that a given condition is met.

  Typically, an automated decision system is "trained" on historical data looking for patterns of relationships between data points (e.g. the relationship between barometer readings, ambient temperature, and snowfall). An automated decision is made by comparing known patterns with existing inputs to estimate how closely they match (e.g. weather prediction based on the similarity between today's climate readings and those from the past). Examples of ADS include algorithms that calculate credit scores and biometric recognition systems that attempt to identify individual people based on physical traits, such as facial features.

- **Automation Bias:** Automation bias is a psychological phenomenon that can occur when operators of an AI system disregard or over-comply with the system's output or are unable to appropriately assess the reliability of its decisions and outcomes for reason of technological prejudice. As such the user can a) become over-reliant on the system and trust it too much, in turn failing to identify inaccurate predictions or classifications, or b) become suspicious of the system and under-use it, despite the fact that it may outperform them on certain tasks.

- **Dataset:** A dataset is a file of information that can typically be represented as a collection of measurements or observations, recorded in a set of rows and columns. Each row corresponds to an individual or an object that can be described using a series of recorded values for each feature that is represented by the series of columns. For example, the following dataset represents a series of measurements for patients at a fictional doctor's surgery, where each patient is provided with a uniquely identifiable patient number:

  | Patient # | Age (Range) | Weight (Kg) | Blood Pressure (mmHG) |
  |---|---|---|---|
  | 1883652 | 26 > 30 | 71 | 115/75 |
  | 1268833 | 31 > 40 | null | 139/83 |
  | 1776436 | 65 > 70 | 90 | 170/90 |
  | 1557821 | 41 > 50 | 72 | 131/82 |

  In the above example, only the first 4 patients are shown, and only 3 features are recorded. However, medical datasets can be vast, not only in terms of the number of patients, but also in terms of the possible values that are recorded. In addition, for patient 1268833 there is no record of their weight. Missing data present a significant challenge for machine learning, and can affect the accuracy of the model that is developed.

- **Equality of arms:** Equality of arms describes the requirements for a person to be subject to a fair trial. To have an equality of arms is expressed in human rights doctrine in the right to an adequate defence, including the right to have access to legal counsel and to call and cross-examine witnesses. Where technologies are used in the conduct of criminal prosecutions, an equality of arms may mean being able to interpret and contest their functions and performance.

- **Explainability:** Closely related to transparency, the explainability of an AI system is the level to which the processes and the rationale behind outcomes of the system can be understood by human users. This can include the extent to which the inner workings of the model can be transformed into plain language, in order to promote better decision-making and trust.

- **Fairness:** Fairness can be defined in many ways, but it can be expressed by the extent to which an AI system promotes or prevents the mitigation of bias and the exclusion of discriminatory influences on its outputs and implementations. Because the AI lifecycle, including the decision to use AI, is affected at every stage by human choices, AI fairness is determined by evaluating human bias and its influence on what AI does and who benefits and does not from its use. In the context of AI, ensuring fairness requires attending to the data employed, the overall design of the system, the outcomes of its use, and decisions about its implementation.

  - *Data fairness* means that data sets used by AI are sufficiently representative of the population likely to be affected, are of high quality and relevance, that the choices that resulted in the data being collected in the first place are examined for bias, and that the data is auditable.

- *Design fairness* means that the activities of system designers are thoughtful, reflective, and mindful of the potential for standpoint bias by the development team. Design fairness requires evaluation of the overall problem formation and chosen outcome, selection and management of data used, feature selection, and whether similar outcomes are achieved for members of different groups and identities. In short, designers should ensure that the systems they produce do not contribute to undesirable social conditions, including harmful discrimination, resource depletion, or oppressive structures of power.

- *Outcome fairness* is an assessment of whether the decisions or other results produced by AI are equitable, fair, and result in distributions of rights, obligations, and public goods in a just manner. Outcome fairness is also an evaluation of the values promoted or prevented by the use of AI.

- We can break down the fairness of an instance of AI as well by the perspectives of the stakeholders who affect or are affected by its use. Each instance of AI has a different and potentially shifting set of stakeholders. General categories include subjects, implementers, and societies.

- To establish *subject fairness*, we can ask if the person who is subject to a decision or action taken by or supported by an AI system perceives the process and outcome as justifiable and legitimate. To establish justifiability and legitimacy, the subject may need to know the details of how the decision or action was arrived at and what factors might have led to another outcome (e.g. a recruiting algorithm rejects a job applicant, which can be explained by showing that the applicant lacks a specifically named skill or credential). A subject also needs access to recourse if they disagree with the outcome (e.g. a job applicant has the opportunity to offer additional information or question the accuracy of the recruiting algorithm to a human with authority to alter the outcome).

- *Implementer fairness* can be expressed through accountability measures including processes of auditing and evaluation. Implementers are tasked with ensuring that AI systems are transparent and interpretable by those who use them and who are affected by that use. Prior to and during the use of AI, implementers should take social, economic, and political effects into account, being mindful not only of the perceived benefits of AI but also for the occurrence and risk of harm and who bears it. For example, the introduction of a criminal sentencing algorithm may produce more judicial consistency and/or streamline decision making. However, the same system may also reproduce discriminatory outcomes, such as where people of colour have received longer sentences for similar convictions as whites in white-majority countries, due to some feature of its design or the data it references. Where such conflicts occur, the functional accuracy or efficiency (if present) of an AI should be set aside and the algorithm design and data model should be thoroughly evaluated, including the decision as to whether to use it.

- *Societal fairness* carries a wider concern. A system whose use has potential impacts on the rights and privileges of individuals, groups, and/or the direction of society requires close attention by human beings and an open deliberative process regarding its use. Policy makers, scholars, and activists are tasked with proposing and critiquing strategies and actions aimed at promoting general well-being and social justice. When AI is used in either private or public sector settings (or both due to public-private partnerships), it potentially participates in preserving or contesting existing social, economic, and political arrangements.

As such, AI should be subject to open and inclusive evaluation for its role in these arrangements, and the humans involved in its design and implementation should be held to account for their choices. Ultimately, the use of AI, like any tool, is acceptable only if it promotes improvements in the conditions of life faced by humans without causing harm.

- **Generalisability:** A model is said to be generalisable when it is effective across a wide range of inputs that reflect real world data, and in a wide range of operational contexts. If a model is not sufficiently trained on representative data it is likely to have limited generalisability when deployed in the real world.

- **Intellectual property:** Intellectual property (IP) describes the products of creative work and their legal possession. Common forms of intellectual property include copyrights, patents, trademarks, and trade secrets. Copyright is a form of IP that protects a creator's right to profit from the authorship of an original work such as a novel, musical composition, or painting. A patent is an exclusive but time-limited licence to profit from the invention and discovery of new and useful processes, machines, articles of manufacture, or compositions of matter. Examples include new medicinal drugs and driverless car technologies. A trademark allows a business entity to reserve the use of a word, name, symbol, or device, or any combination thereof, that identifies its goods and distinguishes them from goods produced by others. An example is the name "Twitter" and associated logos that uniquely identify and distinguish a prominent social media platform. A trade secret is any information that can be used in the operation of a business or other enterprise and that is sufficiently valuable and secret to afford an actual or potential economic advantage over others, such as the recipe for Coca-Cola.

- **Model:** A model is the end result of applying an algorithm to a set of input data (or variables) in order to obtain a predictive or informative output value. Typically, a model is a formal (mathematical) mapping function that aims to represent the underlying processes, and the interactions between them, which are assumed to give rise to relationship between the observed input data and the algorithm's output. For example, the following simple model could express the relationship between a set of input variables, such as the size of a property ($x_1$), the number of bedrooms ($x_2$), the age of the property ($x_3$), and an output variable ($y$), which represents the price. Here, the coefficients or parameters of the x variables are used as weights that signify how important each of the input variables are, based on how much they influence y. The task of the learning algorithm in this case would be to find the values for each parameter that accurately predict the actual house price in the training data. The resulting model could then be used to estimate the prices of new houses, which were not included in the original dataset.

- **Proportionality:** Proportionality is a legal principle that refers to the idea of delivering a just outcome in ways that are proportionate to the cost, complexity, and resources available. In a similar vein, it can also be used as an evaluative notion, such as in the case of a data protection principle that states only personal data that are necessary and adequate for the purposes of the task are collected.

- **Representativeness:** Data used in the algorithm reflects the real world. Does the sample chosen replicate characteristics found in the overall population? An example of non-representativeness is illustrated by the fact that the largest image databases are constructed by people in a small number of countries. A search for "wedding dress" in a typical image database may not identify the marriage attire of many non-western cultures.

- **Socio-Technical System:** A socio-technical system is one that couples human (or social) behaviour to the functionings of a technical system, and in doing so gives rise to novel (and emergent) functions that are not reducible to either the human or technical elements. By intervening in human behaviours, attitudes, or their relations to the world, the technical system restructures human behaviour. The socio-technical perspective is one that considers the human desires or goals a technology is meant to, or does, achieve.

    For example, AI recommender systems that are a common feature of retail, video, and social media sites are socio-technical because they are intended to produce behaviours desired by the operators of the site, such as longer engagement times and/or the purchase of goods. A machine learning algorithm on a video-sharing site analyses the viewing behaviour of thousands or millions of users and makes recommendations to viewers based on their resemblance to a similar subset of users. It is socio-technical both because of its dependence on knowledge about its viewers and because the purpose of its analysis is to keep viewers engaged watching videos, which generates advertising revenue.

    We can also describe as socio-technical those systems whose very existence, implementation, or effects implicate human political, economic, or social relations. For example, surveillance systems adopted by law enforcement agencies are socio-technical because their adoption and use have political dimensions; the selected targets of police surveillance are affected more acutely than others by the use of surveillance technologies based on the historical choices made by government and law enforcement officials. From this socio-technical perspective, surveillance technologies participate in relations between people and the centres of power in society.

- **Soft law:** Soft laws are the policy and regulatory structures that compel or restrain action without the force of state sanctions or penalties. Examples of soft law include 'best practices' and ethics guidelines produced by companies and trade associations. In some professions, such as the practice of law and healthcare, soft law is the set of ethical practices required for certification. Violation of medical ethics can result in a loss of licence to practise medicine. These have varying levels of punitive effect on those subject to them. For example, the Association of Computing Machinery (ACM) has a '[Code of Ethics and Professional Conduct](#)' that is supposed to be followed by its members. However, there are no prescribed sanctions and no system of adjudication for members of the Association who violate the Code. Soft law can also describe the incentive structures reflected in government policy. For example, making tax credits available for producers of 'green' technologies incentivises, but does not compel, production choices.

- **Training/Testing Data:** To build a model and make sure it is accurate, a dataset will typically be split into two smaller sets: training data and testing data. The training data are used to initially develop the model, by feeding the data into an algorithm. Once the model has been trained, it is then tested on the remaining data. The purpose for splitting the data in this manner is to ensure that the model can generalise to new settings, as the data that are collected will only represent a small sample of the overall population. If all the data were used to train the model there is a risk of *overfitting*, which results in a model that performs well for the original dataset but poorly with newer data. Testing a model with "unseen" data also enables data scientists to identify underfitting, i.e. when a model's mapping function fits the data distribution too loosely and is therefore unable to accurately account for the complex patterns it is trying classify or predict.

- **Transparency:** The transparency of AI systems can refer to several features, both of their inner workings and behaviours, as well as the systems and processes that support them. We can describe an AI system as transparent when it is possible to determine how it was designed, developed, and deployed. This can include, among other things, a record of the data that were used to train the system, or the parameters of the model that transforms the input (e.g. an image) into an output (e.g, a description of the objects in the image). However, it can also refer to wider processes, such as whether there are legal barriers that prevent individuals from accessing information that may be necessary to understand fully how the system functions (e.g. intellectual property restrictions).

# APPENDIX 2: COUNCIL OF EUROPE'S AND RELATED WORK IN THE FIELD OF AI AND ADJACENT AREAS TO DATE

This additional reference material has been consolidated from Chapter 4 of the *Feasibility Study*. The number headings correspond to those found in the *Feasibility Study*.

**4.1. Protection of personal data**
- Convention 108/108+ (1981/2018)
    - Processing of sensitive data can only be allowed where appropriate guidelines are present
    - Every individual has the right to know the purpose of processing their data. Along with this, they have a right to rectification and obtainment of knowledge where data are processed contrary to Convention's provisions
    - Transparency, proportionality, accountability, impact assessments, and respect for privacy by design are introduced
    - Individuals should not be subjected to decisions made solely by automated processing of data without consideration of personal views
    - "Legal framework built around Convention remains fully applicable to AI technology, as soon as the processed data fall within the scope of the Convention."
    - Modernised Convention 108+ adopted in 2018; Guidelines on Children's Data Protection in an Educational Setting was adopted by the Convention in November 2020
        - Sets forth "the fundamental principles of children's rights in an education setting and help for legislators and policy makers, data controllers as well as the industry to uphold these rights."

**4.2. Cybercrime**
- Convention on Cybercrime ("Budapest Convention")(2001)
    - "Criminalising offences against and by the means of computers, for procedural powers to investigate cybercrime and secure electronic evidence."
        - Crimes include but are not limited to infringements of copyright, computer-related fraud, child pornography, and violations of a security network
        - Investigation includes a series of powers and procedures including interception and the search of computer networks
    - Primary objective is to "pursue a common criminal policy aimed at the protection of society against cybercrime, especially through appropriate legislation and international co-operation."
    - The cross-border nature of digital networks necessitates a concerted international effort to tackle misuse of technologies
    - Three aims of the convention:
        - "Harmonising the domestic criminal substantive law elements of offences and connected provisions in the area of cyber-crime."
        - "Providing for domestic criminal procedural law powers necessary for the investigation and prosecution of such offences as well as other offences committed by means of a computer system or evidence in relation to which is in electronic form."
        - "Setting up a fast and effective regime of international co-operation."

**4.3. Work in the field of algorithmic systems**
- Declaration on the manipulative capabilities of algorithmic processes (2019)
    - Many individuals are unaware of the dangers of data exploitation
    - Computational means reinforce existing forms of discrimination by sorting individuals into categories
    - The Committee of Ministers draws attention to "the growing threat to the right of human beings to form opinions and take decisions independently of automated systems, which emanates from digital technologies."
    - The primary threats include micro-targeting, identifying vulnerabilities, and the reconfiguration of social environments
    - The Committee gives several recommendations for addressing these threats including but not limited to considering additional protective frameworks that focus on the impacts of targeted use of technologies, initiating open-ended, informed and inclusive public debates about the line between permissible persuasion and unacceptable manipulation, empowering users through increased public awareness and promotion of digital literacy skills, along with several others
- Recommendation on the human rights impacts of algorithmic systems (2020)
    - Member States are advised to review their legislative frameworks, policies, and their own practices to ensure that the procurement, design, and development of algorithmic systems are not violating the human rights framework
    - "Human rights that are often violated through reliance on algorithmic systems include but are not limited to the right to a fair trial; the right to privacy and data protection; the right to freedom of thought, conscience, and religion; the right to freedom of expression; the right to freedom of assembly; the right to equal treatment; and economic and social rights."
    - Additionally, is it recommended that Member States engage in regular, inclusive, and transparent consultation with relevant stakeholders – focusing on the voices of vulnerable groups
    - This recommendation includes various obligations of States with regards to the protection and promotion of human rights and fundamental freedoms in the context of algorithmic systems including obligations such as legislation, transparency, accountability and effective remedies, precautionary measures, etc.
- MSI-AUT Responsibility and AI: A study of the implications of advanced digital technologies (including AI systems) for the concept of responsibility within a human rights framework (2019)
    - This report outlines what AI is and how task-specific technologies work, threats and harms associated with advanced digital technologies, and a range of 'responsibility models' for the adverse impacts of AI systems
    - The main recommendations from this report are "effective and legitimate mechanisms that will prevent and forestall human rights violations", policy choices regarding responsibility models for AI systems, support of technical research involving human rights protections and 'algorithmic auditing', and the presence of legitimate governance mechanisms for the protection of human rights in the digital age
    - Those who develop and implement digital technologies cannot do so without responsibility – they must be held accountable for adverse impacts

**4.4. Work in the field of justice**
- European Ethical Charter on the use of AI in judicial systems and their environment (2018)
  - Five key principles are outlined in this charter including respect for fundamental rights, non-discrimination, quality and security, transparency, impartiality and fairness, and "under user control."
  - Most applications of AI in the judicial field have been found to be in the private sector – "commercial initiatives aimed at insurance companies, legal departments, lawyers, and individuals."
  - Some potential uses of AI in a judicial setting include case-law enhancement, access to law, and the creation of new strategic tools
  - Other considerations that require considerable methodological precautions include the creation of scales, support for alternative dispute settlement measures in civil matters, pre-litigation resolution of disputes online (when a later appeal to the judge remains possible), or identification of where criminal offences are being committed

**4.5. Work in the field of good governance and elections**
- European Committee on Democracy and Governance (CDDG)
  - Currently preparing a study on the impact of digital transformation on democracy and governance
- Venice Commission: Principles for a fundamental rights-compliant use of digital technologies in electoral processes (2020)
  - Emphasised the need for a human rights-compliant approach to eight principles involving the use of digital technologies in elections
  - The eight principles are described in greater detail in the document, but they are outlined below and have been taken directly from the original document
    - 1. "The principles of freedom of expression implying a robust public debate must be translated into the digital environment, in particular during electoral periods."
    - 2. "During electoral campaigns, a competent impartial Electoral Management body (EMB) or judicial body should be empowered to require private companies to remove clearly defined third-party content from the internet – based on electoral laws and in line with international standards."
    - 3. "During electoral periods, the open internet and net neutrality need to be protected."
    - 4. "Personal data need to be effectively protected, particularly during the crucial period of elections."
    - 5. "Electoral integrity must be preserved through periodically reviewed rules and regulations on political advertising and on the responsibility of internet intermediaries."
    - 6. "Electoral integrity should be guaranteed by adapting the specific international regulations to the new technological context and by developing institutional capacities to fight cyberthreats."
    - 7. "The international cooperation framework and public-private cooperation should be strengthened."
    - 8. "The adoption of self-regulatory mechanisms should be promoted."

### 4.6. Work in the field of gender equality and non-discrimination

- [Committee of Ministers Recommendation CM/Rec(2019)1 on preventing and combating sexism](#)
  - The recommendation states that measures must be taken to prevent and combat sexism, along with including a call to integrate gender equality perspective to all work related to AI while finding ways to help eliminate gender gaps and sexism
- [European Commission against Racism and Intolerance (ECRI) - Discrimination, artificial intelligence, and algorithmic decision-making (2018)](#)
  - AI applications have found ways to "escape current laws." Majority of non-discrimination statutes relate only to specific protected characteristics. There are other forms of discrimination that are not correlated with protected characteristics but can still reinforce social inequality
  - The idea of sector-specific rules for the protection of fairness and human rights in the area of AI is proposed, as different sectors necessitate different values and problems
  - For a particular sector, the ECRI proposes several questions that must be answered:
    - "Which rules apply in this sector, and what are the rationales for those rules?"
    - "How is or could AI decision-making be used in this sector, and what are the risks?"
    - "Considering the rationales for the rules in this sector, should the law be improved in the light of AI decision-making?"

### 4.7. Work in the field of education and culture

- [Committee of Ministers' Recommendation CM/Rec(2019)10 on developing and promoting digital citizenship education](#)
  - Invites Member States to adopt regulatory policy measures on digital citizenship education, include all relevant stakeholders in the design, implementation, and evaluation of digital citizenship education legislation, policies and practices, and evaluate the effectiveness of new policies and practices
  - Stresses the importance of "empowering citizens to acquire the skills and competences for a democratic culture, by enabling them to tackle the challenges and risks arising from the digital environment and emerging technologies."
- Steering Committee for Education Policy and Practice (CDPPE)
  - Exploring implications of the use of AI in educational settings
- [Eurimages and the Council of Europe – Entering the new paradigm of artificial intelligence and series (2019)](#)
  - Study on the impact of predictive technologies and AI on the audio-visual sector
  - In this paper, artificial intelligence usage in the audio-visual sector is noted as "a potential threat to the diversity of content and the free access to information of the citizens of the Member States."
  - Five final recommendations are offered ranging from "mandating Eurimages to build competence on Series", "proposing terms of trade for series production in Member States inspired by international best-practice and encourage collaborations" and "raising awareness on the impact of AI in the audio-visual sector."
  - There is also a recommendation that the Council of Europe consider the creation of a "governing body for a media AI certification."

**4.8. Work of the Parliamentary Assembly of the Council of Europe**
- Technological convergence, artificial intelligence, and human rights (2017)
    - Calls for an implementation of "genuine world internet governance that is not dependent on private interest groups or just a handful of States."
    - Additionally, the Assembly calls on the Committee of Ministers to:
        - "Finalise the modernisation of the Convention for the Protection of Individuals with regard to Automatic Processing of Personal Data"
        - "Define a framework for both assistive technologies and care robots in the Council of Europe Disability Strategy 2017-2023."
    - The Assembly also reiterates the importance of accountability and responsibility of AI systems sitting with human beings, informing the public about their personal data generation and data processing that occurs in relation to their personal data, and recognising rights related to respect for private and family life, amongst other proposed guidelines
- 7 reports regarding AI have been adopted by the Parliamentary Assembly with topics ranging from democratic governance to discrimination, and the legal aspects of autonomous vehicles
- Need for democratic governance of artificial intelligence (2020)
    - The Assembly recommends the following:
        - "The elaboration of a legally binding instrument governing artificial intelligence…"
        - "Ensuring that such a legally binding instrument is based on a comprehensive approach, deals with the whole life cycle of AI-based systems, is addressed to all stakeholders, and includes mechanisms to ensure the implementation of this instrument."

**4.9. Work of the Congress of Local and Regional Authorities of the Council of Europe**
- Preparation of "Smart cities: the challenges for democracy" is underway and will be issued in the latter half of 2021

**4.10. Work of the Commissioner for Human Rights**
- Unboxing artificial intelligence: 10 steps to protect human rights (2019)
    - Recommendations are to be used to mitigate or prevent negative impacts of AI systems on human rights
    - Practical recommendations are given with 10 areas of action: human rights impact assessments; public consultations; human rights standards in the private sector; information and transparency; independent monitoring; non-discrimination and equality; data protection and privacy; freedom of expression, freedom of assembly and association, and the right to work; avenues for redress; and promoting knowledge and understanding of AI
    - A checklist is provided to allow for operationalisation of the recommendations contained in the document

**4.11. Work of the Council of Europe in the field of youth**
- Council of Europe Youth Strategy 2030 (2020)
    - Calls for an improvement of institutional responses to emerging issues (including AI) affecting young people's rights and their transition to adulthood
    - The three main focuses of the 2030 strategy are:
        - "Broadening youth participation."
        - "Strengthening young people's access to rights."
        - "Deepening youth knowledge."
    - Additional thematic priorities include increasing capacity for participatory democracy, conducting policies in a way that involves diverse groups of young people, strengthening young people's "capacities, agency, and leadership to prevent violence, transform conflict and to build a culture of peace…", amongst several others

**4.12. Work of the European Committee on Crime Problems (CDPC)**
- [Feasibility study on a future Council of Europe instrument on Artificial Intelligence and Criminal Law (2020)](#)
    - Working group of the CDPC instructed in December 2019 to "carry out a feasibility study identifying the scope and the main elements of a future Council of Europe instrument on AI and criminal law, preferably a convention"
    - Explores the potential of the Council of Europe to pave the way for the adoption of an international legal instrument on AI and criminal law and, on the basis of questionaire replies from member states on AI and criminal law, lays out key elements of an international Council of Europe instrument on AI and criminal law
    - Four objectives of the legal instrument identified:
        i. To establish an international framework for the development of national legislation on criminal law issues in relation to AI (more particularly regarding criminal liability in the context of driving automation);
        ii. To encourage member states to take into account the legal issues in the area of criminal law and AI by addressing problems through legislation, using common normative principles;
        iii. To anticipate the evidentiary and other legal problems already identified in relation to criminal liability and AI and to ensure fair trial-principles as well as effective international co-operation in this area; and
        iv. To ensure the development of AI systems in accordance with the fundamental rights protected by Council of Europe instruments.
    - Study concludes: "agreeing on common standards to clearly and properly allocate possible criminal responsibility and to clarify connected procedural issues as well as possible human rights implication needs to be a joint effort by public and private sector actors, so that the technology can develop successfully and in a way that respects the founding principles of civil society."